\documentclass[10pt]{article} 
\usepackage[utf8]{inputenc}  
\usepackage[english]{babel} 

\usepackage{amsmath}  
\usepackage{amsfonts}        
\usepackage{amssymb}         

\usepackage{booktabs}        
\usepackage{array}           

\numberwithin{equation}{section}    

\usepackage[left=1.50cm, right=1.50cm, top=1.50cm, bottom=1.50cm]{geometry}  
\usepackage{geometry} 
\usepackage{graphicx} 
\usepackage{caption} 

\usepackage{subcaption} 

\usepackage[perpage]{footmisc}

\usepackage{float} 
\usepackage{wrapfig} 

\usepackage{xcolor}
\usepackage{threeparttablex}

\graphicspath{{./Pictures/}} 

\usepackage[titletoc]{appendix} 
\usepackage{appendix}

\usepackage{enumitem} 

\usepackage[nottoc, notlot, notlof, numbib]{tocbibind} 

\usepackage{footmisc} 
\usepackage{hyperref} 
	
\usepackage{lscape} 
\usepackage{longtable} 
\usepackage{multirow} 

\usepackage{multicol} 

\usepackage{cite}  
\let\cite=\citen   

\makeatletter
\renewcommand\@biblabel[1]{}

\makeatother

\begin{document}


\begin{titlepage}

\newcommand{\HRule}{\rule{\linewidth}{0.5mm}} 

\begin{figure}[H] 
\center{\includegraphics[scale=0.35]{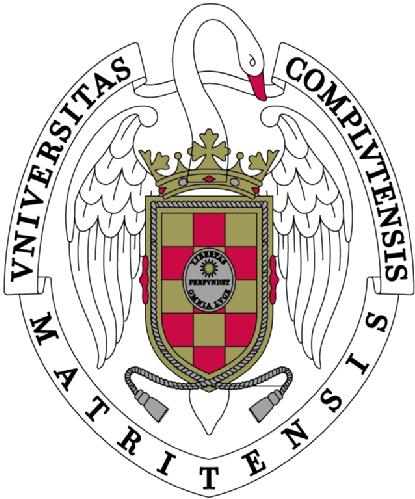}}
\end{figure}

\vspace{1 cm}

\center 

\text{\LARGE Máster Universitario en Astrofísica}\\ \vfill

\text{\LARGE Universidad Complutense de Madrid}\\ \vfill 

\text{\LARGE Trabajo de Fin de Máster}\\ \vfill

\HRule \\[0.4cm]
{ \huge \bfseries Preparación y explotación científica de CARMENES: la metalicidad de las enanas M}\\[0.4cm] 
\HRule \\[1.5cm]

\large
\textbf{Alumno:}\\
Rodrigo \textsc{González Peinado} \footnote{\color{blue}{\underline{rodrgo01@ucm.es}}} 

\vspace{0.5 cm}

\large
\textbf{Directores:} \\
David \textsc{Montes} \footnote{\color{blue}{\underline{dmontes@ucm.es}}\label{David}} (UCM), José Antonio \textsc{Caballero} \footnote{\color{blue}{\underline{caballero@cab.inta-csic.es}}} (LSW)

\vspace{0.5 cm}

\large
\textbf{Tutor:} \\
David \textsc{Montes} \footref{David} (UCM)
  
\vspace{1.5 cm}

{\Large \textbf{Septiembre 2016}}\\[1.5 cm] 

\vfill 

\end{titlepage}

\large
\begin{center}
\textbf{Resumen:}\\
\end{center}
\normalsize

\noindent\textit{Contexto}: CARMENES es un espectrógrafo de alta resolución con el que el consorcio hispano-alemán del mismo nombre busca exotierras alrededor de unas 300 estrellas enanas de tipo espectral M por el método de velocidad radial.

\noindent\textit{Objetivos}: Recopilar información de utilidad en diferentes catálogos para una muestra inicial dada de 209 sistemas estelares binarios y múltiples, formados por una primaria de tipo espectral F, G o K y una compañera secundaria de tipo M (o K tardía). Comprobar si el par de estrellas forma efectivamente un par de movimiento propio y obtener calibraciones de metalicidad espectroscópicas y fotométricas en banda $K$ a partir de estos sistemas binarios.

\noindent\textit{Métodos}: La recopilación de datos de cada estrella se ha realizado buscando en catálogos especializados en VizieR y la bibliografía. La comprobación de si las estrellas son compañeras de movimiento propio se ha realizado estudiando los movimientos propios de ambas estrellas con la ayuda de dos herramientas del observatorio virtual: $Aladin$ y $TopCat$. A partir de una lista de sistemas aptos, se han obtenido dos tipos de calibraciones de metalicidad: espectroscópicas y fotométricas. Para derivar dichas calibraciones, se ha supuesto que la metalicidad de la primaria, calculada por el grupo de investigación de la UCM dedicado a CARMENES, es igual a la de la secundaria.

\noindent\textit{Resultados}: Se ha obtenido una calibración espectroscópica dependiente únicamente de la anchura equivalente de Na~{\sc i} (2.206 $\mu$m y 2.209 $\mu$m) y una fotométrica que mejora las publicadas hasta ahora. A su vez, se han estimado metalicidades para 134 estrellas M en sistemas múltiples, con un rango de metalicidad --0.9$<$[Fe/H]$<$+0.4.

\noindent\textit{Conclusiones}: El estudio de los sistemas lejanos y físicos de esta muestra ha permitido diseñar calibraciones de metalicidad espectroscópicas y fotométricas en la banda $K$ que podrán usarse para calcular metalicidades de estrellas M aisladas.\\
\

\noindent\textbf{Palabras clave:} Bases de datos astronómicas --- Estrellas: movimientos propios  --- Estrellas: tipo tardío  --- Estrellas: compañeras --- Estrellas: metalicidad ---


\vspace{1 cm}

\large
\begin{center}
\textbf{Abstract:}\\
\end{center}
\normalsize

\noindent\textit{Context}: CARMENES is a next-generation instrument being built by a consortium of German and Spanish institutions to carry out a survey of 300 M-type dwarf stars with the goal of detecting exoearths by radial-velocity measurements.

\noindent\textit{Aims}: To collect relevant information from different on-line catalogues for a given sample of 209 binary or multiple star systems, formed by F, G or K primary star and an M-dwarf (or late-K) companion. To prove if the pair is indeed a physical pair, to obtain different metallicity calibrations in $K$-band with these binary systems.

\noindent\textit{Methods}: The data compilation from every star has been done searching in catalogues in VizieR and the literature. In addition, physical pair checking has been done studying the collected proper motions from both stars (primary and secondary) and using two tools from the Virtual Observatory: $Aladin$ and $TopCat$. From a list of suitable systems, two different types of calibrations had been obtained: spectroscopic and photometric. In order to determine these calibrations, we have considered that metallicity from the primary star, determined by the CARMENES UCM research group, is equal to the secondary star.

\noindent\textit{Results}: The spectroscopic calibration obtained is only dependant on the Na~{\sc i} (2.206 $\mu$m y 2.209 $\mu$m) equivalent width and the photometric calibration determined here improves the ones published so far. In addition, metallicities had been obtained for 134 M-dwarfs in multiple systems, in a range of -0.9$<$[Fe/H]$<$+0.4.

\noindent\textit{Conclusions}: The study of wide binary physical systems in our sample allows to derive spectroscopic and photometric metallicity callibrations in $K$-band that could be used to calculate metallicities in isolated M-dwarfs. \\
\



\newpage 



\renewcommand{\contentsname}{Índice}

\tableofcontents 

\newpage 


\section{Introducción} \label{Introduction}
\subsection{CARMENES}
\label{CARMENES}
CARMENES \footnote{\url{https://carmenes.caha.es/}} (Calar Alto high-Resolution search for M-dwarfs with Exoearths with Near-infrared and optical Échelle Spectrographs) es un espectrógrafo échelle de alta resolución situado en el telescopio Zeiss de 3.5 m del Observatorio de Calar Alto, a 2168 metros sobre el nivel del mar en la Sierra de los Filabres, Almería (España). El Instituto de Astrofísica de Andalucía (IAA) junto con el Max-Planck-Institut für Astronomie (MPIA) son los encargados de operar el Centro Astronómico Hispano-Alemán (CAHA), donde se sitúa CARMENES (Quirrenbach et al. 2014).\\

\noindent Como consorcio, CARMENES está formado por 11 instituciones: MPIA, IAA, el Landessternwarte Königsthul (LSW) de Heidelberg, el Institut de Ciències de l'Espai (ICE) de Barcelona, el Institut für Astrophysik Göttingen (IAG), la Universidad Complutense de Madrid (UCM), el Thüringer Landessternwarte Tautenburg (TLS), el Instituto de Astrofísica de Canarias (IAC), el Hamburger Sternwarte (HS), el Centro de Astrobiología (CAB) en Madrid y el CAHA. Más de 150 científicos e ingenieros de estas instituciones han participado en el desarrollo, construcción y explotación científica de este nuevo "cazador" de planetas.\\

\noindent Lo que hace a CARMENES único y diferente de otros espectrógrafos échelle de alta resolución dedicados a la búsqueda de exoplanetas, es su capacidad para observar en dos regímenes de longitud de onda diferentes. Así, CARMENES es un espectrógrafo de dos canales, uno en el visible (que cubre un rango de longitud de onda entre 520 y 960 nm en 55 órdenes) y otro en el infrarrojo cercano (cuyo rango cubre entre 960 y 1710 nm en 28 órdenes). Posee una resolución de 94600 en el visible y de 80400 en el infrarrojo cercano (NIR). La calibración en longitud de onda se lleva a cabo mediante un sistema de lámparas de emisión (U-Ar, U-Ne y Th-Ne), además de un \textit{étalon} de Fabry-Pérot para cada canal. La temperatura de trabajo en el visible es de 285,00$\pm$0,05\,K, mientras que en el infrarrojo cercano necesita ser refrigerado hasta los 140,00$\pm$0,05\,K para alcanzar una precisión en velocidad radial de 1 m/s.\\

\noindent La técnica en la que CARMENES se basa para la detección de exoplanetas se conoce como velocidad radial (RV, del acrónimo en inglés). Este método da cuenta del efecto Doppler que se produce cuando una estrella y un objeto que la orbita, en este caso un planeta, se mueven en torno al centro de masas común del sistema. CARMENES está optimizado para operar en el infrarrojo cercano y posee una precisión de 1 m/s, lo que le permitirá encontrar planetas dentro de la zona de habitabilidad de las estrellas M. Se ampliará más sobre este tema en la Sección \ref{enanasM}.\\

\noindent Con motivo de la búsqueda de exoplanetas alrededor de este tipo de estrellas, se creó la base de datos Carmencita (CARMENes Cool dwarf Information and daTa Archive; Caballero et al. 2013 y Caballero et al. 2016). Carmencita posee aproximadamente unas 2200 enanas M y un gran número de parámetros para cada objeto, bien extraídos de la literatura o bien medidos por el consorcio. Entre los parámetros destacan la información astrométrica y cinemática de la enana ($\alpha$, $\delta$, $\mu_{\alpha}cos\delta$, $\mu_{\delta}$, $\pi$, $V_{r}$, $U$, $V$, $W$), tipos espectrales (desde M0.0V hasta M9.5V) e información fotométrica (19 bandas fotométricas cubriendo un rango de longitudes de onda desde el ultravioleta al infrarrojo). La obtención de otros parámetros como metalicidades, gravedad superficial, temperatura o indicadores de actividad y edad ($pEW$(H$\alpha$), información de rayos X, velocidad rotacional) ha servido para seguir ampliando la información en Carmencita y poder descartar aquellas enanas no adecuadas para la búsqueda de exoplanetas con el método de la velocidad radial debido a su alta actividad, baja gravedad superficial, binarias espectroscópicas y rotadoras rápidas. Otro de los aspectos importantes de Carmencita es el tratamiento de la multiplicidad de los sistemas. Respecto a esto, se ha dedicado la Sección \ref{1.3} para su explicación.
Para un mejor aprovechamiento del tiempo de observación de CARMENES, se han seleccionado las $\sim$300 estrellas M más prometedoras de Carmencita. Estas $\sim$300 finalistas son los objetivos de observación durante las 600 noches hasta 2018 dedicadas al proyecto. Todas ellas cumplen las siguientes tres características, además de las expuestas anteriormente: 
\begin{itemize}
\item Deben ser observables desde Calar Alto , es decir, poseer una declinación~$\delta$ $>$ --23 deg (distancia cenital $z$ $<$ 60\,deg, masa de aire en culminación $<$ 2.0) 
\item Deben ser las estrellas M más brillantes en cada subtipo espectral.
\item Deben ser estrellas individuales sin compañera a una distancia mínima angular $\rho$ $<$5 arcsec.
\end{itemize}

\noindent Así, Carmencita es el catálogo más completo de enanas M hasta la fecha, accesible a todos los investigadores que forman parte del consorcio CARMENES y que se espera que sea público en un futuro, conformando uno de los legados de CARMENES.\\

\noindent Por todo esto, CARMENES se convierte en uno de los referentes mundiales presente y futuro de búsqueda de exoplanetas y una prueba clara de la necesidad de colaboración entre distintos países e instituciones a la hora de hacer ciencia.\\

\subsection{Enanas M y búsqueda de exoplanetas}
\label{enanasM}
Las enanas M son los objetos estelares más comunes en la Vía Láctea, al menos en la vecindad solar. De hecho, la estrella más cercana al Sol, Proxima Centauri, es una enana M (M5.5V, $V$=11.23 mag). Debido a que son objetos sumamente fríos y pequeños, su temperatura oscila entre los 3800\,K para las más calientes y 2300\,K para las más frías, no pueden ser localizados a simple vista y se precisa de telescopios para verlos. Las enanas M poseen espectros característicos poblados por bandas moleculares en el óptico (Reid \& Hawley 2005). Entre las bandas moleculares más comunes, predominan las producidas por óxido de titanio (TiO) en absorción y en todo el espectro (las más prominentes se dan en 6322, 6569, 6651, 7053, 7666, 8206 y 8432 \AA). Tambień destacan las absorciones de hidruro de calcio (CaH) en 6346, 6382 y 6750 \AA\ y de óxido de vanadio (VO) en las M de últimos tipos en torno a 7334 y 7851~\AA. En cuanto a características atómicas, son notables la línea de H$\alpha$ 6563 \AA\ en emisión en M tardías, así como el doblete de potasio en 7665--7699 \AA\ y el doblete de sodio en 8183--8195 \AA. El estudio de los espectros de estos objetos fue lo que llevó a Kirkpatrick et al. (1991) a generar la clasificación espectral que actualmente se toma como estándar. Los autores reanalizaron los estudios anteriores de Boeshaar (1976) y Boeshaar \& Tyson (1985) sobre enanas M y extendieron el régimen de estudio utilizando longitudes de onda mayores (6300--9000 \AA), generando una clasificación de enanas M en diez subtipos, desde M0V (las más calientes) a M9V (las más frías), pero véase también la clasificación más moderna de Alonso-Floriano et al. (2015 a).\\

\noindent La distinción de varios subtipos de enanas M caracteriza muy bien a la estrella, pues cada subtipo lleva asociado unas características propias. Entre los parámetros que describen a las enanas M destacan su temperatura efectiva (\textit{T$_{\rm eff}$}), el radio ($R$), la masa ($M$), la luminosidad ($L$) y la gravedad superficial ($g$). La Tabla \ref{Tabla 1}, obtenida de Reid \& Hawley (2005), muestra cómo varían estos parámetros con el tipo espectral.

\begin{table}[H]
\renewcommand{\tablename}{Tabla}
\caption{Propiedades fundamentales de las enanas M.}
\label{Tabla 1}
\begin{center}
\begin{tabular}{c c c c c c}
\hline
\hline
\noalign{\smallskip}
Spectral & \textit{T$_{\rm eff}$} & $R$ & $M$ & $L$ & log $g$\\
Subtype & [K] & [$R_{\odot}$] & [$M_{\odot}$] & [10$^{-2}L_{\odot}$] & [c.g.s]\\

\noalign{\smallskip}
\hline
\noalign{\smallskip}
M0 & 3800 & 0.62 & 0.60 & 7.2 & 4.65\\
M1 & 3600 & 0.49 & 0.49 & 3.5 & 4.75\\
M2 & 3400 & 0.44 & 0.44 & 2.3 & 4.8\\
M3 & 3250 & 0.39 & 0.36 & 1.5 & 4.8\\
M4 & 3100 & 0.36 & 0.20 & 0.55 & 4.9\\
M5 & 2800 & 0.20 & 0.14 & 0.22 & 5.0\\
M6 & 2600 & 0.15 & 0.10 & 0.09 & 5.1\\
M7 & 2500 & 0.12 & $\sim0.09$ & 0.05 & 5.2\\
M8 & 2400 & 0.11 & $\sim0.08$ & 0.03 & 5.2\\
M9 & 2300 & 0.08 & $\sim0.075$ & 0.015 & 5.4\\
\noalign{\smallskip}
\hline

\end{tabular}
\end{center}
\end{table}

\noindent Las enanas M juegan un papel determinante en la búsqueda de exoplanetas, y más aún si el objetivo es encontrar planetas parecidos a la Tierra. Como se ha explicado en la Sección \ref{CARMENES}, el objetivo de CARMENES es la búsqueda de planetas que orbiten alrededor de este tipo de estrellas por el método de la velocidad radial. Aunque la búsqueda de exoplanetas se comenzó en el régimen óptico por motivos de instrumentación con el descubrimiento del primer exoplaneta en torno a una estrella de tipo solar, nombrado 51 Peg b (Mayor \& Queloz 1995), en 2006 se demostró que el futuro descubrimiento de exoplanetas en torno a estrellas M sería más efectivo en el infrarrojo cercano con el estudio de la enana marrón LP 944-20 (M9V; Martín et al. 2006). \\
Las ventajas de la búsqueda de exoplanetas que orbiten alrededor de enanas M en el infrarrojo cercano son numerosas. La Tabla \ref{Tabla 1} muestra alguna de ellas. En primer lugar, su baja temperatura hace que su emisión sea más intensa en esta longitud de onda. Si a su baja temperatura le sumamos también el hecho de que son poco masivas, esto supone una gran ventaja a la hora de utilizar el método de la velocidad radial anteriormente descrito. Una temperatura baja se traduce en medidas de RV menos afectadas por la actividad estelar, mejorando el estudio de la curva de velocidad radial. A su vez, si la estrella es poco masiva, como en este caso, el movimiento de la estrella en torno al centro de masas del sistema será más notable, por lo que se facilita el descubrimiento de un objeto compañero que puede ser un exoplaneta. Por último, el hecho de que las enanas sean objetos fríos también hace que la zona de habitabilidad de la estrella, es decir, la zona en la que es posible la existencia de agua líquida, esté más próxima y la probabilidad de encontrar planetas como la Tierra aumente. Aún así, hay que tener cuidado con estos estudios, pues los falsos positivos de exoplanetas son frecuentes debido a la contaminación de la curva de velocidad radial por la actividad cromosférica y/o rotación de la propia estrella, que introducen cierta variabilidad que puede enmascarar al exoplaneta. Recientemente, un planeta de masa terrestre (1.3 M$_{\oplus}$) ha sido descubierto orbitando la anteriormente citada Próxima Centauri gracias al método de la velocidad radial, con un perido orbital de 11.2 días (Anglada-Escudé et al. 2016).\\
Se han realizado numerosos estudios de búsquedas de exoplanetas en el infrarrojo cercano, gracias al gran número de espectrógrafos infrarrojos que se han construido en los últimos años, entre los que destacan CRIRES (Seifahrt \& Käufl 2008), Giano (Carleo et al. 2016) o CSHELL (Gagné et al. 2016). La ventaja que posee CARMENES frente a todos los anteriores es que cuenta con dos espectrógrafos, uno en infrarrojo cercano y otro en el visible, que le conceden una precisión en RV de 1 m/s, capaz de detectar exotierras de 2 M$_{\oplus}$ en la zona de habitabilidad de una enana de tipo M5V. Por eso, esperamos que CARMENES se convierta en el espectrógrafo de referencia para la búsqueda de exoplanetas en el hemisferio norte en los próximos años. Mención también a otros espectrógrafos que están actualmente en construcción, como IRD (Tamura et al. 2012), HPF (Mahadevan et al. 2014) o SPIRou (Donati et al. 2014) y futuros, como HIRES (Maiolino et al. 2013), NIRES (Skidmore 2015) o GMTNIRS (Lee et al. 2010), los cuales se esperan que estén operativos para finales de la próxima década en los telescopios de próxima generación (30-40 m.).\\

\subsection{Sistemas, multiplicidad y movimientos propios}
\label{1.3}
\begin{figure}[H]
\centering
\renewcommand{\figurename}{Figura}
\begin{subfigure}{1.1\textwidth}
  \centering
  \includegraphics[width=1\linewidth]{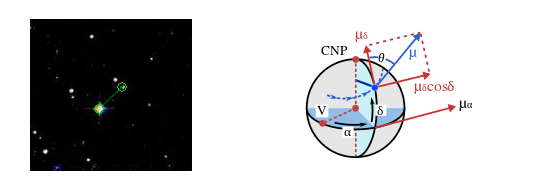}
\end{subfigure}
\caption {{\small $Izquierda$: sistema WDS 03150+0101, prototipo de sistema doble (\textit{The Binary Star Database}); $Derecha$: componentes del movimiento propio, en rojo, y movimiento propio total, en azul. CNP: acrónimo en inglés de Polo Norte Celeste. (Wikipedia)}}
\label{fig1}
\end{figure}

El $Modern$ $Dictionary$ $of$ $Astronomy$ $and$ $Space$ $Technology$ (Bhatia 2005) define un sistema estelar como un pequeño número de estrellas que orbitan entre sí y que están ligadas gravitatoriamente (en adelante, sistemas físicos). Si el sistema está compuesto por dos estrellas, se le conoce como sistema binario y es estable si no existen fuerzas externas ni transferencia de masa de una estrella a otra. En este caso, el sistema orbitará en torno al centro de masas común en órbitas elípticas. Un ejemplo de sistema binario puede verse en la Figura \ref{fig1}. Si el número de estrellas en el sistema es mayor o igual que tres, el sistema se conoce como múltiple. Así, la multiplicidad del sistema indica el número de componentes estelares resueltas que lo conforman. Debido a que estamos situados dentro de la Vía Láctea, la visión de las estrellas está sesgada y la definición de multiplicidad debe ser matizada. Puede ocurrir que se observen dos estrellas aparentemente muy juntas en el cielo, pero que en realidad no forman un sistema físico debido a que una componente está a una distancia muy diferente de la otra, aunque ambas se sitúen en la misma línea de visión. A este tipo de sistemas se los denomina ópticos y deben eliminarse de la muestra si el estudio se centra en sistema estelares físicos.\\

\noindent Una forma de discernir si un sistema es físico o no, es decir, si está ligado y las componentes coevolucionan, es estudiando sus movimientos propios. Como es bien sabido, el movimiento propio de las estrellas hace referencia al cambio de posición angular de éstas en el cielo medido desde el Sol. Este movimiento se define a partir de la suma vectorial de dos componentes en un sistema de coordenadas ecuatorial: la componente es ascensión recta $\mu_{\alpha}$ y la componente en declinación $\mu_{\delta}$. La unidad utilizada para medir movimientos propios es el arcosegundo por año. Como el arcosegundo es una unidad demasiado grande en este caso, suele utilizarse un submúltiplo, el miliarcosegundo por año (mas a$^{-1}$), para representar los movimientos propios. Con el satélite Gaia acabamos de entrar en el régimen del mas a$^{-1}$, lo que supone una revolución en la medida de los movimientos propios debido a su gran precisión. La suma cuadrática de ambas componentes da el movimiento propio total de la estrella:

\begin{equation}
\label{ec1}
\mu=\sqrt{(\mu_\alpha \cos{\delta})^{2}+(\mu_{\delta})^{2}}
\end{equation}

\noindent en donde el factor $\cos{\delta}$ se introduce para dar cuenta del radio de la esfera celeste, pues éste varía con el coseno de la declinación $\delta$. La Figura \ref{fig1} muestra una representación de las componentes del movimiento propio y el movimiento propio total.\\

\noindent Para determinar si un sistema estelar es físico u óptico a partir de los movimientos propios, se comparan tanto el movimiento propio en ascensión recta como en declinación, además del total. Si el movimiento propio en ascensión recta (declinación) de una estrella del sistema no es compatible en un intervalo determinado con el correspondiente de la otra estrella, el par puede calificarse como óptico, pues no estarían gravitacionalmente ligados y su evolución habría sido y sería distinta. Se ampliará el tratamiento de movimientos propios para discernir entre sistemas físicos de no físicos en las Secciones \ref{OV} y \ref{comprobacion}\\

\noindent Dentro de los sistemas binarios, existen un tipo especial denominados sistemas binarios lejanos (\textit{wide binaries}).  La separación mínima entre dos estrellas para que el sistema binario se considere lejano no está bien definida, ya que algunos autores consideran una separación entre componentes igual al tamaño típico de los núcleos protoestelares de los que nacen las estrellas, es decir, 0.1 pc, mientras que otros establecen que la separación mínima depende de la masa del sistema (Caballero 2010). Los sistemas binarios lejanos son de gran utilidad, pues sirven para determinar propiedades de la materia oscura si los sistemas se encuentran en el halo de la galaxia o para el estudio de modelos de formación estelar. La aplicación que se estudiará en este trabajo será utilizar estos sistemas como calibradores de metalicidad para estrellas de tipo M. En la Sección \ref{Metalicidad} se ampliará información sobre este método.\\ 

\subsection{Metalicidad}
\label{Metalicidad}
Uno de los parámetros más importantes para caracterizar una estrella es su metalicidad. La metalicidad se define como la proporción (en logaritmo) entre el número de átomos por unidad de volumen de cualquier elemento en una estrella frente a la de hidrógeno respecto a la misma proporción en el Sol, esto es:

\begin{equation}
[X/H]=\log_{10}\left(\dfrac{N_{X}}{N_{H}}\right)_{*}-\log_{10}\left(\dfrac{N_{X}}{N_{H}}\right)_{\odot}
\end{equation}

\noindent donde $N_{X}$ y $N_{H}$ son el número de átomos por unidad de volumen del elemento problema y el hidrógeno, respectivamente. La unidad que se usa habitualmente para la metalicidad es el $dex$ (contracción de exponente decimal), de modo que una [X/H] negativa significa una proporción del elemento X inferior a la solar y una [X/H] significa una proporción del elemento X superior a la solar. Usualmente, la metalicidad se define a partir del hierro. Aunque no es el elemento pesado más abundante, sí que es de los más fáciles de medir en el espectro debido al gran número de líneas. Por ello, a partir de ahora y siempre que se haga referencia a la metalicidad, se considerará la proporción entre átomos por unidad de volumen de hierro frente a los de hidrógeno respecto al Sol, es decir, [Fe/H]. También es habitual referirse a la metalicidad general, esto es, la proporción de todos los elementos frente al hidrógeno respecto al Sol, que se denota con [M/H].\\

\noindent La medida de la metalicidad es crucial en el descubrimiento de exoplanetas. Estudios en estrellas de tipo F, G y K han demostrado que la presencia de exoplanetas aumenta con la metalicidad de la estrella huésped (Valenti \& Fischer 2005). Como se ha explicado en la Sección \ref{enanasM}, los espectros de las enanas M están muy poblados de absorciones moleculares. Esto supone que en sus espectros sea muy difícil medir una metalicidad fiable. Por ello, se buscan calibraciones con sistemas binarios o múltiples en los que se conoce bien la metalicidad de la componente primaria (FGK) y se asume igual para la(s) otra(s) componente(s) más frías tipo M (Johnson \& Apps 2009; Rojas-Ayala et al. 2012; Newton et al. 2014). La hipótesis de que las metalicidades pueden considerarse iguales para ambas componentes se debe a la suposición de que las dos o más estrellas del sistema físico se han formado en la misma nube molecular, por lo que las componentes presentan la misma (o muy parecida) composición química, y no ha existido transferencia de materia ni dragado de elementos químicos (Rojas-Ayala et al. 2010). Los estudios de esta índole suelen consistir en una muestra de sistemas binarios formados por una estrella primaria de tipo espectral F, G o K (brillante) y una enana de tipo espectral M, separadas por mínimo 5 arcosegundos. La composición de la primaria es determinada utilizando técnicas de espectroscopía (en general, de alta resolución) mientras que para la secundaria se asume la misma composición que para la primaria. Así, si se consigue una expresión o un modelo empírico que permita conectar varios parámetros mensurables de las componentes, se podrá utilizar para calcular la metalicidad de estrellas M aisladas.\\

\noindent Existen principalmente dos tipos de calibraciones de metalicidad: fotométricas y espectroscópicas. En las calibraciones fotométricas se recopilan datos de fotometría de ambas componentes y se relacionan con la metalicidad. Estas calibraciones fueron primero realizadas por Bonfils et al. (2005), que establecieron una relación empírica entre el plano fotométrico ($V-K_{S}$, $M_{K}$) y la metalicidad. Johnson \& Apps (2009) refinaron dicha calibración y obtuvieron una relación lineal entre la metalicidad y la diferencia entre la magnitud absoluta en banda $K$ para la secuencia principal de la vecindad solar y la observada. Siguiendo con este método, Schlaufman \& Laughlin (2010) y Neves et al. (2012) obtuvieron relaciones lineales muy similares entre ellas entre la metalicidad y la diferencia entre los colores $V-K_{S}$ observados y de la secuencia principal de la vecindad solar, consiguiendo las mejores calibraciones fotométricas en enanas M hasta la fecha.\\

\noindent Por otro lado, las calibraciones espectroscópicas se basan en la medida de anchuras equivalentes de varios elementos en el espectro de las estrellas M. El primer estudio de este tipo fue llevado a cabo por Rojas-Ayala et al. (2010), que midieron la anchura equivalente del doblete de Na~{\sc i} (2.206 $\mu$m y 2.209 $\mu$m) y el triplete de Ca~{\sc i} (2.261 $\mu$m, 2.263 $\mu$m y 2.265 $\mu$m) en banda $K$, además del índice de absorción de H$_{2}$O para obtener una expresión para la metalicidad. La misma técnica siguieron Terrien et al. (2012) y Rojas-Ayala et al. (2012), obteniendo expresiones similares. Newton et al. (2014) obtuvo una expresión cuadrática para la metalicidad que sólo dependía de la anchura equivalente de Na~{\sc i}, simplificando notablemente el tratamiento. Actualmente, las calibraciones en el infrarrojo cercano con el mayor número de objetos ($\sim$112) son Mann et al. (2013) para enanas M de primeros y medianos tipos y Mann et al. (2014) para tipos tardíos, utilizando la misma técnica que los grupos anteriores. Estas calibraciones de Mann ofrecen un rango amplio de metalicidades (--0.8 $<$[Fe/H]$<$ 0.5) y tipos espectrales (últimas K8 a M8) con gran precisión ($\sim0.10$ dex en [Fe/H] y [M/H]).

\newpage

\subsection{Objetivos}
\noindent Esta memoria se ha realizado en estrecha colaboración con el grupo de investigación de la Universidad Complutense de Madrid dedicado a CARMENES. Así, el objetivo de este Trabajo Fin de Máster es doble:
\begin{itemize}
\item En primer lugar, se analizará una lista de sistemas estelares proporcionada por dicho grupo de investigación, en la que se estudiará si los pares son físicos o no a partir del estudio de sus movimientos propios. Este paso pretende descartar los sistemas ópticos (y estrellas no válidas, tales como binarias espectroscópicas SB2 o rotadores rápidos) y quedarse con una lista de sistemas físicos lo más amplia posible.
\item En segundo lugar, a partir de la lista ya limpia de sistemas ópticos y estrellas no aptas, se estudiarán diferentes calibraciones de metalicidad, tanto espectroscópicas como fotométricas, con el objetivo de realizar unas calibraciones propias y comprobar si son compatibles con aquellas publicadas en la literatura anteriormente nombradas en la Sección \ref{Metalicidad}.
\end{itemize}

\noindent Para lograr estos dos objetivos, me serviré de herramientas del observatorio virtual como $Aladin$ o $TopCat$, así como de diferentes catálogos accesibles desde VizieR.

\newpage

\section{Recopilación de datos y análisis} \label{Analysis}
\subsection{Muestra inicial y tipos espectrales}
\noindent La muestra inicial consta de 209 pares de estrellas, formados por una estrella primaria de tipo espectral F, G o K, y una secundaria de tipo espectral M (o, en algunos casos, K tardío), recogidos de la literatura utilizando los criterios anteriormente descritos sobre separación, tipo espectral, etcétera. Parte de la muestra puede verse en la Tabla \ref{Tabla 2}. En ella se muestra el número de identificación del sistema (Identificador WDS) y el código del descubridor, obtenidos ambos del catálogo WDS (Mason et al. 2001), la separación entre las componentes ($\rho$), el ángulo de posición ($\theta$), el nombre de las componentes que aparece en Simbad y las coordenadas de las estrellas ($\alpha$ y $\delta$), además del tipo espectral. Se ampliará esta información en la Sección \ref{OV}.\\

\begin{table}[H]
\renewcommand{\tablename}{Tabla}
\caption{Muestra inicial de pares.}
\label{Tabla 2}
\begin{center}
\begin{tabular}{l c c c l c c c c}
\hline
\hline
\noalign{\smallskip}
Identificador & Código & $\rho$ & $\theta$ & Nombre & $\alpha$ & $\delta$ & Tipo \\
WDS & descubridor & [arcsec] & [deg] & Simbad & (J2000) & (J2000) & espectral \\

\noalign{\smallskip}
\hline
\noalign{\smallskip}
00153+5304 &  &  &  & G 217-41 & 00:15:14.8 & +53:04:27 & K3\\
 & GIC 5 & 18.8 & 355 & G 217-40 & 00:15:14.6 & +53:04:46 & M2.5V\\
 \noalign{\smallskip}
00385+4300 & & & & BD+42 126  & 00:38:29.2  & +43:00:00 & G5\\
 & LDS 5176 & 53.1 & 125 & LP 193-345 & 00:38:33.2 & +42:59:30 & M0.5V\\
 \noalign{\smallskip}
00452+0015 &  &  &  & HD 4271 A & 00:45:11.0 & +00:15:12 & F8\\
 & LDS 836 & 55.4 & 45 & HD 4271 B & 00:45:13.6 & +00:15:51 & M4.0V\\
 \noalign{\smallskip}

\noalign{\smallskip}
\hline

\end{tabular}
\begin{tablenotes}
      \small
      \item Nota: esta es una versión reducida de la muestra. La tabla completa puede verse en el Apéndice \ref{aped.B} (Tabla \ref{muestracompleta}).\\
    \end{tablenotes}
\end{center}
\end{table}

\noindent El objetivo será el de proporcionar la máxima información para cada sistema y para cada componente individual. En total, la muestra inicial consta de 201 primarias FGK y 200 secundarias M. La diferencia en el número entre primarias, secundarias y pares totales es debido a que algunos de los sistemas de la muestra inicial poseen una multiplicidad mayor que dos. Este hecho se tratará en la Sección \ref{comprobacion}.\\

\begin{figure}[H]
\centering
\renewcommand{\figurename}{Figura}
\begin{subfigure}{1\textwidth}
  \centering
  \includegraphics[width=1\linewidth]{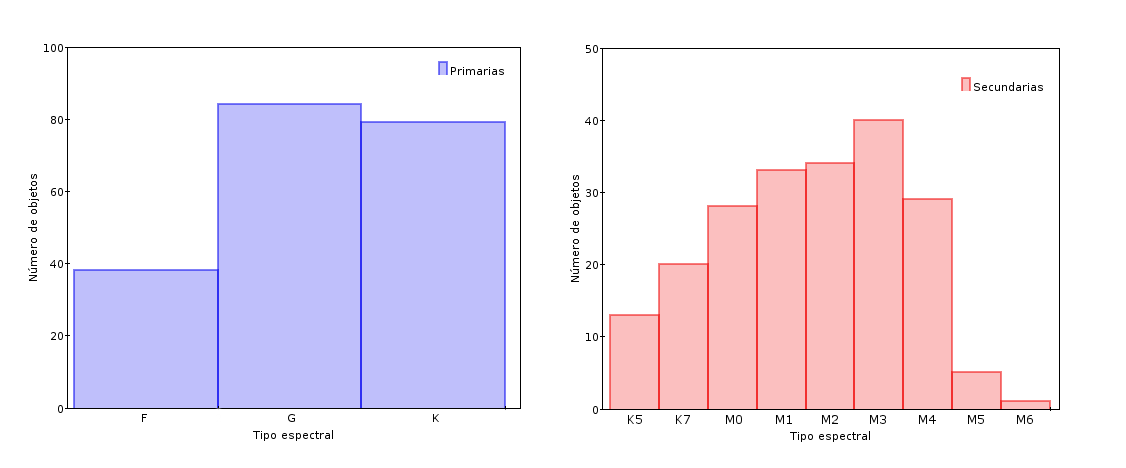}
\end{subfigure}
\caption {{\small Histograma de tipos espectrales en primarias (izquierda) y secundarias (derecha)}.}
\label{fig2}
\end{figure}

\noindent En cuanto a los tipos espectrales, se han recopilado 201 tipos espectrales de primarias provenientes de las bases de datos Simbad y/o VizieR. De las 201 primarias, los tipos G y K representan aproximadamente el 81\% de la muestra inicial (84 primarias tipo G y 79 tipo K) mientras que tan sólo el 19\% son de tipo F (38). 
Por otro lado, para secundarias se han recopilado 193 tipos espectrales, la mayoría provenientes de Alonso-Floriano et al. (2015 a). De éstas, un 16\% son de tipo K tardías (K5-7V, 31 estrellas), mientras que la gran mayoría (78\%) corresponde a tipos M tempranos ($\leq$ M4V, 156 estrellas) y sólamente un 3\% (6 estrellas) son M tardías ($>$ M5). Del 4\,\% de estrellas restante (7), listadas en la Tabla \ref{Tabla 3}, no se han podido recopilar tipo espectral de la bibliografía. Sin embargo, se han estimado sus tipos espectrales a partir de sus colores $r'-J$, calculados a partir de magnitudes $r'$ del Sloan Digital Sky Survey DR9 y Carlsberg Meridian Telescope 14 y $J$ de 2MASS (y a partir de la diferencia de magnitud $J$ entre primaria y secundaria en el caso del par cercano WDS~10504--1326). Todas las secundarias son enanas M tempranas excepto la estrella G de fondo en el sistema WDS 14575--2125. En la Figura \ref{fig2} se puede ver la distribución en tipos espectrales en forma de histograma.   

\begin{table}[H]
\renewcommand{\tablename}{Tabla}
\caption{Pares con secundarias sin tipo espectral espectroscópico.}
\label{Tabla 3}
\begin{center}
\begin{tabular}{l c c c l c c c c}
\hline
\hline
\noalign{\smallskip}
Identificador & Código & $\rho$ & $\theta$ & Nombre & $\alpha$ & $\delta$ & Tipo \\
WDS & descubridor & [arcsec] & [deg] & & (J2000) & (J2000) & espectral \\

\noalign{\smallskip}
\hline
\noalign{\smallskip}
09393+1319 &  &  &  & HD 83509 & 09:39:17.2 & +13:18:45 & F7V \\
 & TOK 270 & 50.6 & 132.0 & J09391981+1318118 & 09:39:19.8 & +13:18:12 & m1:\\
 \noalign{\smallskip}
10504-1326 &  &  &  & BD-12 3277 & 10:50:22.4 & -13:26:07 & K0V\\
 & LDS4023 & 8.6 & 20 & LP 731-61 & 10:50:22.7 & -13:26:00 & m:\\
 \noalign{\smallskip}
12051+1933 &  &  &  & BD+20 2678 A & 12:05:07.0 & +19:33:16 & G5V\\
 & GIC 103 & 117.3 & 144 & BD+20 2678 B & 12:05:11.9 & +19:31:41 & m2: \\
 \noalign{\smallskip}
13076-1415 &  &  &  & HD 114001 & 13:07:39.2 & -14:11:17 & F5V\\
 & TOK 286 & 63.3 & 208 & J13073714-1412130 & 13:07:37.1 & -14:12:13 & m3:\\
 \noalign{\smallskip}
14575-2125 &  &  &  & HD 131977 & 14:57:28.0 & -21:24:56 & K4V\\
 & H N 28 & 198.3 & 323 & GSC 06180-00916 & 14:57:19.5 & -21:22:16 & g:\\
 \noalign{\smallskip}
21546-0318 &  &  &  & HD 208177 & 21:54:35.9 & -03:18:05 & F5IV\\
 & STF2838 & 15.7 & 182 & BD-03 5329 B & 21:54:35.6 & -03:18:18 & m0:\\

\noalign{\smallskip}
\hline

\end{tabular}
\end{center}
\end{table}

\subsection{Observatorio virtual y catálogos}
\label{OV}

\noindent El primer paso para la caracterización y estudio de los 209 sistemas es la observación de los mismos. Para ello, se ha utilizado la herramienta del observatorio virtual $Aladin$. Este programa permite ver cualquier elemento del cielo introduciendo sus coordenadas y la posibilidad de cargar varios catálogos de información del objeto introducido. El proceso empleado para la visualización de los sistemas ha sido el siguiente:
\begin{enumerate}
\item Se carga la imagen astronómica. En este caso, se utilizará el $survey$ proporcionado por la ESO. Para ello, se sigue la siguiente ruta: \textit{File $>$ Load astronomical image $>$ DSS... $>$ DSS from ESO (Garching/Deutschland-DSS.ESO)}. De esta manera, se despliega el panel de la Figura \ref{fig3}.
\begin{figure}[H]
\centering
\renewcommand{\figurename}{Figura}
\begin{subfigure}{0.6\textwidth}
  \centering
  \includegraphics[width=1.2\linewidth]{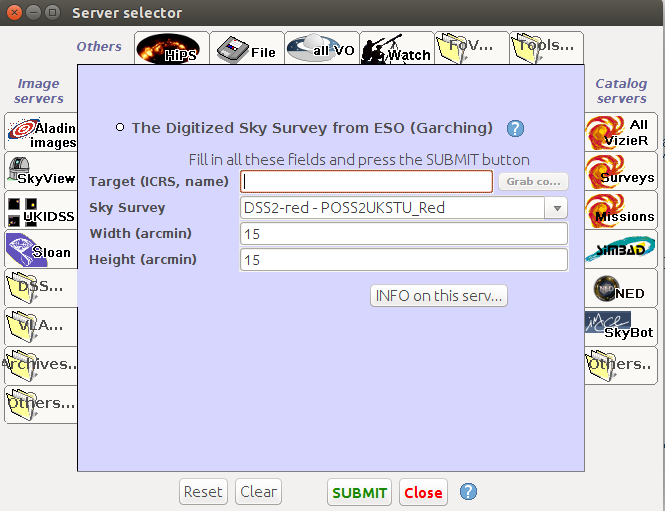}
\end{subfigure}
\caption {{\small Panel de carga de imágenes astronómicas del $survey$ de la ESO en $Aladin$}.}
\label{fig3}
\end{figure}
\item En el renglón de entrada $Target$ $(ICRS, name)$ se introducen las coordenadas de la estrella primaria. Es preferible realizar la búsqueda por coordenadas, pues la búsqueda por nombre puede producir errores. Una vez se han introducido las coordenadas, se pulsa {\em SUBMIT}. Desde este panel también se puede cambiar el tamaño de la imagen y el $survey$ utilizado. De momento, se trabajará con los que aparecen por defecto.
\item Para la carga de información, es necesario fijarse en la parte derecha del panel. En primer lugar, se carga la información de la base de datos Simbad. Para ello, se pulsa el logo de Simbad y se presiona {\em SUBMIT}. De nuevo, desde el panel se puede ajustar el tamaño de la imagen y si se quiere que únicamente se muestre un tipo de objeto determinado (estrellas, galaxias, fuentes de rayos X...). 
\item A continuación, se carga el catálogo WDS (\textit{Washington Double Star Catalogue}, Mason et al. 2001) que proporciona información sobre sistemas binarios y múltiples. Es el catálogo de referencia para el estudio de multiplicidades. Para ello, en la pestaña $Surveys$ se introduce {\em B/WDS} y se pulsa {\em SUBMIT}.\\
\end{enumerate}

\begin{figure}[H]
\centering
\renewcommand{\figurename}{Figura}
\begin{subfigure}{1\textwidth}
  \centering
  \includegraphics[width=0.95\linewidth]{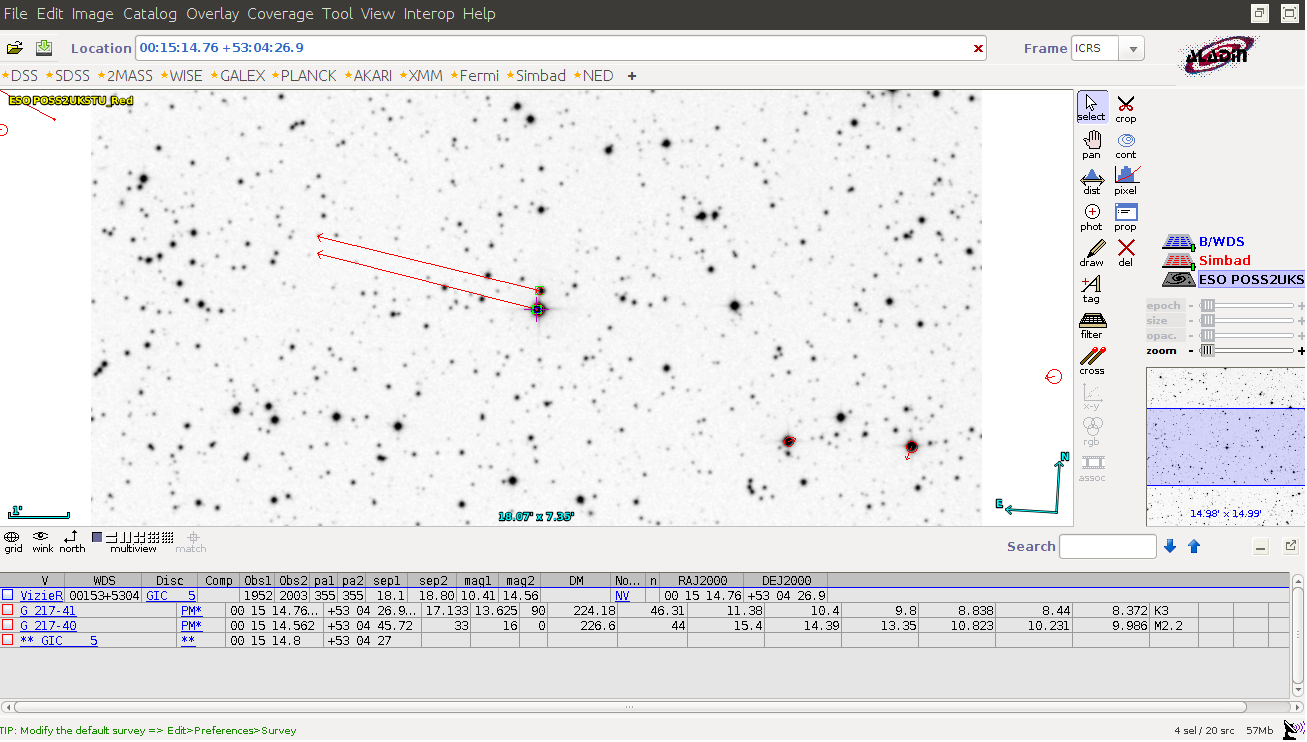}
\end{subfigure}
\caption {{\small Sistema WDS 00153+5304 visualizado desde $Aladin$}.}
\label{fig4}
\end{figure}

\noindent A modo de ejemplo, se pueden seguir los pasos descritos previamente con el sistema WDS 00153+5304 de la Tabla \ref{Tabla 2} del que únicamente se conocen las coordenadas de sus componentes. Introduciendo las coordenadas de la estrella primaria y siguiendo los pasos explicados, se despliega en la pantalla de $Aladin$ una imagen centrada en la primaria con objetos marcados. Si se selecciona con el ratón el sistema central, aparece información sobre ambas estrellas en la parte inferior de la pantalla. Es importante tener en cuenta la orientación de la imagen, pues el Norte apunta hacia arriba y el Este, a la izquierda. Esto ayudará a identificar mejor algunas características del sistema, como su separación angular ($\rho$) y, sobre todo, el ángulo de posición ($\theta$). En la Figura \ref{fig4} se muestra la imagen final que se obtiene para el sistema WDS 00153+5304.

\noindent El código de colores del panel de la derecha es el que se sigue en el de abajo. Así, el color rojo representa la información en Simbad y el azul, la información en WDS. Del catálogo WDS extraemos el identificador WDS, el código del descubridor $Disc$ (tres o cuatro letras más un número), la separación entre componentes $\rho$ (que en $Aladin$ figura como $sep$. Se toma la medida más reciente, en arcosegundos) y el ángulo de posición $\theta$ entre ellas ($pa$ en $Aladin$. De nuevo, la más reciente).\\ Para comprobar que se selecciona el sistema adecuado, se puede realizar el siguiente procedimiento. Con la herramienta $dist$ de la parte derecha de la Figura \ref{fig4} se traza una línea recta desde la primaria hasta la que se considere que es la secundaria. Esta medida tiene que coincidir con alguno de los valores de separación en el caso de que haya más de dos componentes en el sistema. Para asegurarse todavía más de que el sistema es el correcto, se puede medir sobre la pantalla el ángulo de posición teniendo en cuenta la orientación de la imagen y teniendo en cuenta que el ángulo de posición se mide siempre de Este a Oeste partiendo desde la línea que une la estrella primaria con el Norte. Otro aspecto importante del catálogo WDS es el apartado de notas ($Notes$). Aquí se presentan diferentes anotaciones y correcciones sobre los sistemas y que pueden ayudar en caso de duda.\\

\noindent Una vez que ya se ha extraído la información deseada del catálogo WDS, se pasa a recopilar otros datos de interés para la caracterización de la muestra, como movimientos propios, paralajes o magnitudes fotométricas de ambas componentes. A continuación se hace un repaso de los catálogos más importantes utilizados para cada medida:

\begin{itemize}

\item Movimientos propios: $Aladin$ permite visualizar los movimientos propios sobre las estrellas con varios catálogos. Para ello, se cargan desde la pestaña $Surveys$ del panel de la Figura \ref{fig3}. Algunos catálogos de referencia en cuanto a movimientos propios son \textit{Hipparcos, the New Reduction} (van Leeuwen, 2007), \textit{The Tycho-2 Catalogue} (H$\o$g et al. 2000), \textit{UCAC4 Catalogue} (Zacharias et al. 2012) o \textit{The PPMXL Catalog} (Roeser et al. 2010). La recopilación de movimientos propios se ha realizado utilizando este mismo orden. Estos datos serán los que se usarán para discernir si un sistema es físico y por lo tanto válido o si por el contrario es óptico y es necesario desecharlo.

\item Paralajes: la gran mayoría de paralajes se han obtenido del catálogo de $Hipparcos$. Para aquellas estrellas para las que $Hipparcos$ no poseía paralaje, se ha acudido a la base de datos VizieR y se ha recopilado la paralaje con referencia más reciente.  Se ha calculado la distancia para cada componente a partir de la relación entre ésta y la paralaje $\pi$ de la siguiente forma:

\begin{equation}
d\,\text{[pc]}=\dfrac{1}{\pi\text{[arcsec]}}
\end{equation}

\item Magnitudes fotométricas: se han recopilado las magnitudes en las bandas $V$, $J$, $H$ y $K_{S}$ para cada componente del sistema del catálogo \textit{2MASS All-Sky Catalog of Point Sources} (Skrutskie et al. 2006) y del catálogo \textit{UCAC4 Catalogue} (Zacharias et al. 2012). En el caso de que no existiese entrada para la estrella, se han utilizado los catálogos \textit{GSC2.3} (Lasker et al. 2008), \textit{From binaries to multiples. I. The FG-67 sample} (Tokovinin 2014) o \textit{Revised NLTT Catalog} (Salim \& Gould 2003) en este orden, todos ellos accesibles desde VizieR.
\end{itemize}

\begin{figure}[H]
\centering
\renewcommand{\figurename}{Figura}
\begin{subfigure}{0.82\textwidth}
  \centering
  \includegraphics[width=0.95\linewidth]{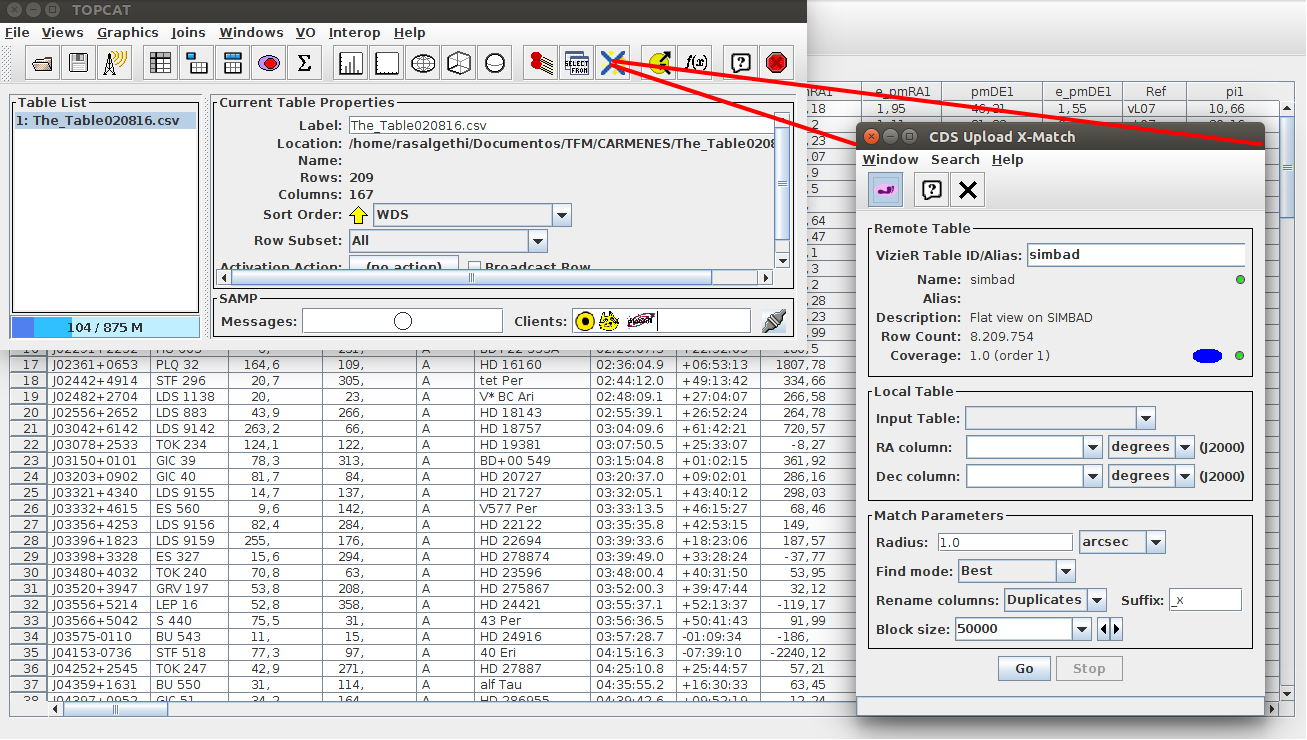}
\end{subfigure}
\caption {{\small Interfaz de $TopCat$. \textit{Arriba izquierda}: panel de control. \textit{Derecha}: ventana de $crossmatch$. \textit{Fondo}: tabla de trabajo en formato .csv.}}
\label{fig5}
\end{figure}

\noindent Todos estos datos han sido recopilados en forma de tabla en formato .csv accesible desde la herramienta del observatorio virtual $TopCat$. Para automatizar el proceso de toma de datos, desde $TopCat$ se puede realizar un $crossmatch$ con los catálogos presentes en VizieR. Para ello, sólo es necesario pulsar el botón \textit{Sky crossmatch against remote tables} en el panel de contro de $TopCat$ e introducir el código del catálogo deseado además de las columnas de ascensión recta y declinación para identificar el objeto en dicho catálogo, como se muestra en la Figura \ref{fig5}. Este proceso devuelve tantas columnas de información como entradas tenga el catálogo, lo cual reduce el tiempo de trabajo.\\

\noindent En el Apéndice \ref{aped.B}, al final de la memoria, se adjuntan los valores de astrometría y fotometría recopilados (Tablas \ref{astrometry} y \ref{photometry}).\\

\subsection{Comprobación de sistemas por movimientos propios}
\label{comprobacion}
\begin{figure}[H]
\centering
\renewcommand{\figurename}{Figura}
\begin{subfigure}{0.95\textwidth}
  \centering
  \includegraphics[width=0.95\linewidth]{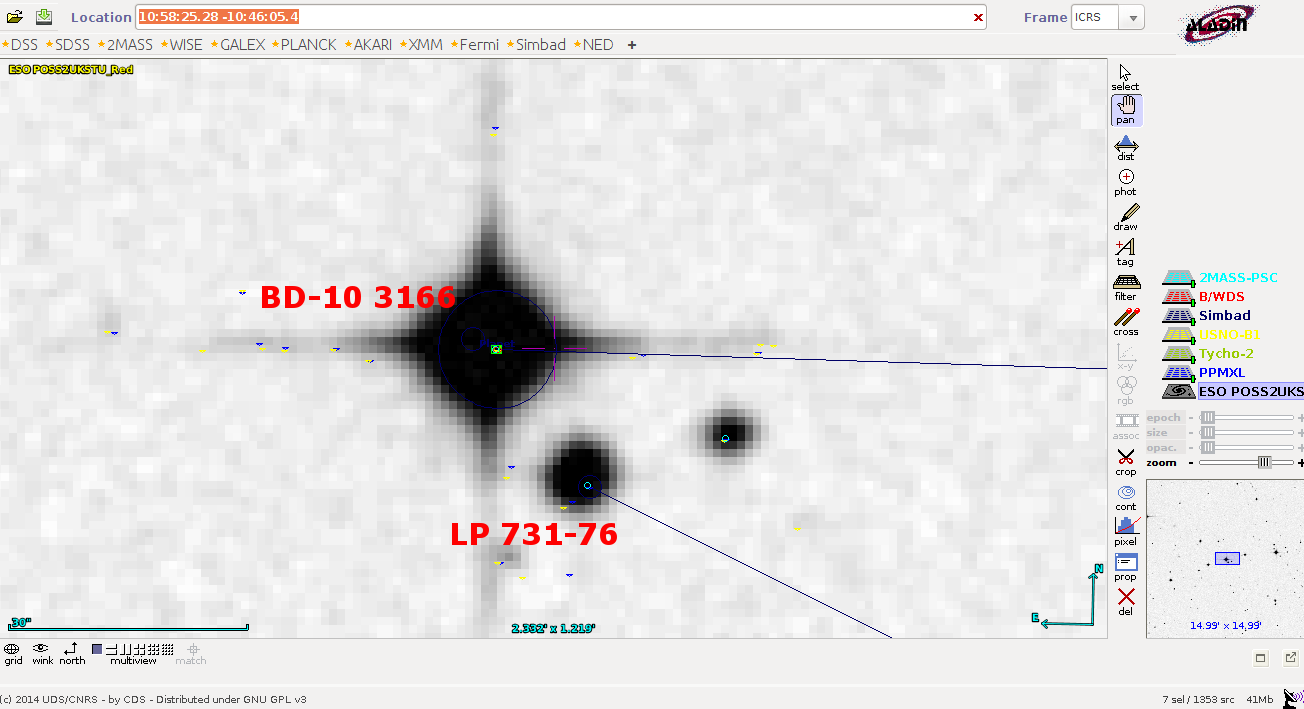}
\end{subfigure}
\caption {{\small Imagen de $Aladin$ del sistema WDS 10585-1046 formado por la primaria BD-10 3166 y la secundaria LP 731-76. Se aprecia como claramente las líneas de movimientos propios no son paralelas. Por tanto, el sistema no es físico y deberá ser descartado si los movimientos propios están bien medidos.}}
\label{figsistema}
\end{figure}

\noindent Existen varias maneras de discernir si un sistema es físico o no. Para una primera inspección, se puede representar el movimiento propio en ascensión recta y en declinación de la estrella primaria frente a la secundaria. En el caso de que las dos componentes formasen parte del mismo sistema, los puntos se dispondrían a lo largo de una recta. Si por el contrario no formasen parte del mismo sistema, cada componente tendría un movimiento propio diferente y caerían fuera de dicha recta. Las Figuras \ref{fig6} y \ref{fig7}, adjuntadas en el Apéndice \ref{aped.A}, ilustran este hecho. Como se ve, existen varios puntos que no siguen la tendencia general. Estos puntos serán los que habría que estudiar para comprobar si son o no sistemas físicos. Para automatizar el procedicimiento, se ha adoptado el criterio de selección utilizado en Alonso-Floriano et al. (2015 b). Este criterio considera como compañeros físicos aquellos sistemas cuya diferencia relativa de $\mu_\alpha \cos{\delta}$ y $\mu_{\delta}$ de la secundaria están dentro de un 10\% del valor correspondiente de la primaria. Para la componente de ascensión recta se tendría:

\begin{equation}
\label{ecra}
\dfrac{\Delta\mu_{\alpha} \cos\delta}{\overline{\mu_{\alpha} \cos\delta}}=\Big\vert\dfrac{\mu_\alpha \cos{\delta}_{B}-\mu_\alpha \cos{\delta}_{A}}{\overline{\mu_\alpha \cos{\delta}}}\Big\vert<10\%
\end{equation}
\\
\noindent donde los subíndices A y B se refieren a la componente primaria y secundaria respectivamente. En el caso de la componente en declinación sería:

\begin{equation}
\dfrac{\Delta\mu_{\delta}}{\overline{\mu_{\delta}}}=\Big\vert\dfrac{\mu_{\delta B}-\mu_{\delta A}}{\overline{\mu_{\delta}}}\Big\vert<10\%
\end{equation}
\\
\noindent A su vez, también se ha realizado el mismo procedimiento para el movimiento propio total. En este caso, el criterio es el siguiente:

\begin{equation}
\label{ectot}
\dfrac{\Delta\mu}{\mu_{A}}={\left[{\dfrac{(\mu_\alpha \cos{\delta}_{A}-\mu_\alpha \cos{\delta}_{B})^{2}+(\mu_{\delta A}-\mu_{\delta B})^{2}}{\mu_\alpha \cos{\delta}_{A}^{2}+\mu_{\delta A}^{2}}}\right]}^{1/2}<10\%
\end{equation}
\\
\noindent Este criterio no tiene en cuenta qué ocurre cuando, por ejemplo, las componentes del movimiento propio en ascensión recta cumplen la condición establecida en \ref{ecra}, pero no lo hacen las componentes en declinación o el total. Esto es frecuente en aquellos casos en los que alguna de las componentes del movimiento sea muy pequeña, pues su 10\% podría no estar dentro del rango de valores de la primaria. Este caso, en el que una o las dos componentes del movimiento propio sea muy pequeñas en comparación con la otra ($<$ 10 mas a$^{-1}$) se han tomado como compañeras siempre y cuando el criterio establecido en la Ecuación \ref{ectot} sea razonable.\\
Para los casos en los que los movimientos propios no sean pequeños pero no se cumplan los criterios arriba establecidos, se ha realizado la siguiente comprobación para descartarlos o incluirlos definitivamente en la muestra final. El procedimiento consiste en cargar en $Aladin$ varios catálogos de movimientos propios como \textit{Hipparcos} (van Leeuwen 2007),\textit{ Tycho-2} (H$\o$g et al. 2000), PPMXL (Roeser et al. 2010), NOMAD (Zacharias et al. 2005), UCAC4 (Zacharias et al. 2012)... desde el panel de carga de catálogos. Así, obtenemos una medida de la misma componente del movimiento propio pero en años diferentes. De esta se forma realiza una segunda comprobación para que ningún par sea desechado injustificadamente. Un ejemplo de esto se muestra en la Figura \ref{figsistema}. El sistema representado es WDS 10585-1046, uno de los sitemas de la muestra.\\

\medskip

\newpage
\noindent Además de descartar sistemas por movimientos propios, es necesario realizar una segunda inspección para eliminar aquellos sistemas no válidos para las calibraciones. Los siguientes criterios aplican tanto a una estrella individual (primaria o secundaria) o al sistema completo, por lo que si cualquiera de los objetos del sistema, o el sistema en sí, presenta las siguientes característas, deberán ser descartados:
\begin{itemize}
\item Sistemas con separación angular $\rho$ entre componentes menor de 5 arcsec: la presencia de una componente a una distancia angular menor de 5 arcsec puede introducir errores en la magnitud de la estrella objetivo o contaminar su espectro, perjudicando así la realización de las calibraciones. Por eso es necesario establecer una separación angular mínima en la que ambas componentes no interfieran mutuamente. Con este criterio se descartan los sistemas visuales y/u ópticos, quedando únicamente sistemas lejanos (\textit{wide binaries}).
\item Binarias espectroscópicas SB2: es un caso partícular del caso anterior. Las binarias espectroscópicas son sistemas que, aunque no pueden ser resueltos por medio de telescopios, el estudio de sus espectros puede revelar la presencia de dos estrellas. En particular, en las binarias espectroscópicas SB2, la luminosidad de la secundaria no es despreciable y puede contaminar tanto la fotometría como la espectroscopía. Por esta razón, se excluirán todas aquellas estrellas, tanto primarias o secundarias, que sean SB2. Por otro lado, en las binarias espectroscópicas SB1, una de las componentes es tan débil que no interfiere en la medida de la fotometría y/o espectroscopía. Por ello, las binarias espectroscópicas SB1 no han sido excluídas de la muestra.
\item Movimiento propio total de la primaria $\mu<$ 50 mas a$^{-1}$: este criterio permite eliminar aquellos sistemas lentos de movimiento, ya que podrían no ser separables del $\mu$ del resto de objetos de fondo y, por tanto, no permitiría distinguir si las dos componentes son o no físicas.
\item Estrella rotadoras rápidas: cuando una estrella rota muy rápido, en torno a $v\sin i\sim10-12$ km/s, se produce un ensanchamiento de las líneas espectrales. Este fenómeno no permite determinar correctamente las anchuras equivalentes de las líneas para determinar los parámetros estelares y las abundancias. Estas estrellas se han obtenido aplicando el código $StePar$ (Tabernero et al. 2012 y 2013) sobre la muestra inicial y, posteriormente, han sido descartadas.
\item Primarias con \textit{T$_{\rm eff}>$} 6500 K: las estrellas calientes tienen el inconveniente de poseer muy pocas líneas en sus espectros y, generalmente, estar ensanchadas para la medida de anchuras equivalentes. Por ello, se han descartado aquellas estrellas cuyo tipo espectral es $<$F4V, que es el límite de aplicación de $StePar$.
\item Secundarias con magnitud en banda $J$ $<$ 15.5 mag: objetos tan débiles en el infrarrojo cercano lo serían aún más en el óptico, dificultando así la medida de movimientos propios. De esta manera, mantener objetos más brillantes de 15.5 mag en banda $J$ asegura buenas medidas de movimientos.
\end{itemize}

\noindent La aplicación de estos criterios de selección permitirá obtener una muestra limpia, a partir de la cual se podrán realizar las calibraciones de metalicidad.

\subsection{Parámetros de metalicidad}
\label{metpar}
\noindent El otro área de estudio de esta memoria se refiere a las metalicidades de los sistemas. Como se explicó en la Seccion~\ref{Metalicidad}, existen dos formas de estudio de la metalicidad: espectroscopía y fotometría. A continuación se describen ambos métodos y los valores seleccionados para su estudio.
\subsubsection{Metalicidad de primarias FGK}
\label{metprim}
\noindent Como se ha descrito en la Sección \ref{Metalicidad}, las calibraciones de metalicidad se realizan suponiendo para ambas componentes la misma metalicidad. Al ser más fácil medir la metalicidad de las primarias, se toma ésta como metalicidad del sistema. En el grupo de investigación de la UCM dedicado a CARMENES, y, por extensión, en este trabajo, las metalicidades de las primarias se han obtenido mediante la aplicación de $StePar$ (Tabernero et al. 2013), que toma los espectros de alta resolución de las primarias y mide las anchuras equivalentes de líneas de Fe para determinar parámetros estelares (\textit{T$_{\rm eff}$}, $\log g$, $\xi$, [Fe/H]) y, a partir de ellos, calcular la abundancia de Fe y otros elementos utilizando el código MOOG. Estos espectros han sido obtenidos con el espectrógrafo HERMES (R$\sim$85000; Raskin et al. 2011) situado en el telescopio Mercator de La Palma. Así, las metalicidades se han determinado de una forma homogénea para todas las primarias FGK fente a los valores utilizados por otros autores. En adelante, a los valores de metalicidad de las primarias obtenidos mediante $StePar$ se nombrarán como [Fe/H]$^{*}$.\\
\subsubsection{Anchuras equivalentes en NIR de las secundarias M}
\noindent El principal elemento de estudio en las calibraciones espectroscópicas son las anchuras equivalentes de ciertas líneas absorción sobre el espectro de una estrella. Para este trabajo se han recopilado las anchuras equivalentes correspondientes al doblete de Na~{\sc i} (2.206 $\mu$m y 2.209 $\mu$m) y al triplete de Ca~{\sc i} (2.261 $\mu$m, 2.263 $\mu$m y 2.265 $\mu$m) en banda $K$ de estrellas M. A su vez, también ha sido necesario recopilar el índice H$_{2}$O-K, que da cuenta de la influencia de la temperatura en la profundidad de las líneas. Se ha dividido el tratamiento de esta Sección en dos muestras:
\begin{itemize}
\item Muestra 1: anchuras equivalentes de ambas especies e índices H$_{2}$O-K recopilados de Newton et al. (2014), Rojas-Ayala et al. (2012) y Rojas-Ayala et al. (2010) por este orden de preferencia. Con estos datos se han calculado valores de [Fe/H] según las calibraciones de Rojas-Ayala et al. (2012), Terrien et al. (2012) y Newton et al. (2014). El propósito de este cálculo será comparar las metalicidades obtenidas para secundarias con las deducidas por el grupo de investigación dedicado a CARMENES (ver más abajo) mediante la realización de un ajuste lineal.
\item Muestra 2: anchuras equivalentes del doblete de Na~{\sc i} recopiladas de Mann et al. (2015), Mann et al. (2014), Mann et al. (2013) y Newton et al. (2014) por este orden de preferencia. Con estos datos se pretende realizar una calibración propia de metalicidad dependiente exclusivamente de la anchura equivalente de Na~{\sc i}.
\end{itemize} 
\noindent Ambas muestras pueden encontrarse en el Apéndice \ref{aped.B}, al final de este trabajo (Tablas \ref{muestra1} y \ref{muestra2}).

\subsubsection{Magnitudes fotométricas de las secundarias M}
\label{magfot}

\noindent El objetivo de este apartado es obtener una expresión que relacione la metalicidad de la estrella primaria con las características fotométricas de la secundaria. Para ello, se seguirán los mismos pasos que los explicados en Schlaufman \& Laughlin (2010) y Neves et al. (2012). El estudio se realizará sobre el plano \{\textit{$V-K_{S}$, M$_{K_S}$}\}. Se asumirá una relación lineal entre la metalicidad y las características fotométricas de las estrellas secundarias tal que [Fe/H]$\propto\Delta(V-K_S)$, en donde $\Delta(V-K_S)=(V-K_S)_{obs}-(V-K_S)_{iso}$ es la distancia hasta el contorno (o traza) de isometalicidad de la muestra en el plano \{\textit{$V-K_{S}$, M$_{K_S}$}\}. Para generar un contorno de isometalicidad los más preciso posible, se han seleccionado aquellas primarias cuyo valor de metalicidad se sitúa en un intervalo 2$\sigma$ respecto a la metalicidad solar (Tabla \ref{sigma2}, Apéndice \ref{aped.B}). Este tratamiento supone una diferencia con los estudios fotométricos mencionados en la Sección~\ref{Metalicidad}, pues éstos sólo limitan su traza de isometalicidad a la vecindad solar, generando un contorno de isometalicidad sesgado a estrellas cercanas, mientras que en este trabajo, el contorno de isometalicidad se realizará con estrellas de metalicidad solar independientemente de su distancia, por lo que será más representativo y no se limitará tan solo al entorno solar.\\

\noindent Se seleccionarán únicamente aquellas secundarias que presenten magnitud en banda $V$ de UCAC4 para conseguir una fotometría lo más homogénea posible, mientras que la magnitud en banda $K_{S}$ se ha obtenido de 2MASS. Las estrellas secundarias para el ajuste fotométrico se presentan en el Apéndice \ref{aped.B} (Tabla \ref{ucac_ajuste}). Con estas magnitudes, se reproducirán las calibraciones de Bonfils et al. (2005), Johnson \& Apps (2009), Schlaufman \& Laughlin (2010) y Neves et al. (2012). Estas calibraciones se realizan a partir de la magnitud absoluta de la estrella. Para derivar la magnitud absoluta de las estrellas de la muestra, se hará uso del módulo de distancias, el cual permite obtener la magnitud absoluta a partir de la magnitud relativa y la distancia a la estrella, es decir:

\begin{equation}
m_{\lambda}-M_{\lambda}=5\log d\ -5
\end{equation}

\noindent siendo $m_{\lambda}$ y $M_{\lambda}$ las magnitudes relativa y absoluta en la banda $\lambda$ respectivamente y $d$, la distancia a la estrella en parsecs. Se ha escogido la distancia a la estrella primaria para el cálculo de la magnitud absoluta, pues es más precisa debido a que está medida sobre una estrella más brillante que su compañera secundaria.\\

\section{Resultados y discusión} \label{Results_discussion}
\noindent En esta sección se recogen los resultados obtenidos dividos en dos partes. En primer lugar, se presentarán los resultados  derivados de la observación de cada sistema en $Aladin$ y se expondrán aquellos sistemas que no eran sistemas físicos o se han descartado por otras razones. En segundo lugar, se presentarán los resultados propios de metalicidades, diferenciando entre espectroscópicas y fotométricas, obteniéndose sendas calibraciones. Para el desarrollo de este apartado, se han utilizado las herramientas del observatorio virtual $Aladin$ y $TopCat$.
\subsection{Multiplicidad y movimientos propios}
\noindent En primer lugar se mostrarán los sistemas con multiplicidad mayor que dos para después centrarse en aquellos que serán descartados, exponiendo las razones para ello. Por último, se presentará la muestra final de estudio a partir de la cual se realizarán las calibraciones de metalicidad.
\subsubsection{Multiplicidad de los sistemas}
\label{multiplicity}
\begin{figure}[H]
\centering
\renewcommand{\figurename}{Figura}
\begin{subfigure}{0.95\textwidth}
  \centering
  \includegraphics[width=0.95\linewidth]{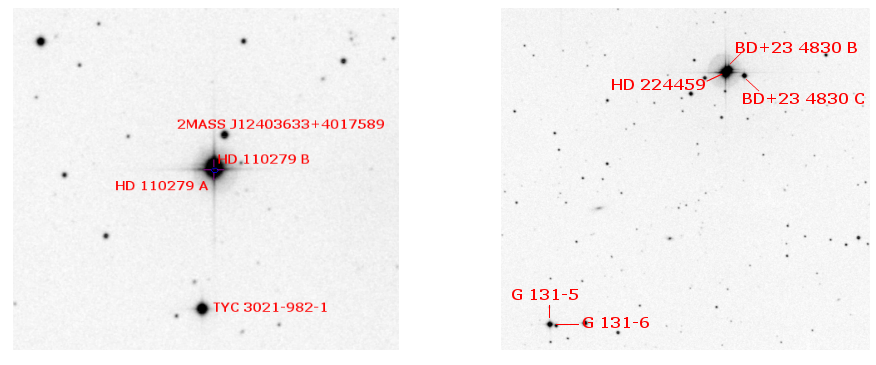}
\end{subfigure}
\caption {{\small Imagen de $Aladin$ de los sistemas WDS 12406+4017 (izquierda) y WDS 23581+2420 (derecha).}}
\label{dossistemas}
\end{figure}

\noindent En la muestra inicial se han encontrado diez sistemas múltiples, los cuales se describen a continuación (no se muestran aquí binarias visuales próximas):
\begin{itemize}
\item WDS 05445-2227: sistema triple conformado por dos primarias: $\gamma$ Lep (tipo espectral F6V) y AK Lep (K2V); y la secundaria LHS 1781 (M3.5V). Ambos sistemas son ópticos y, por tanto, han sido descartados.
\item WDS 07041+7514: sistema triple conformado por dos primarias: HD 50281 A (G0) y HD 50281 B (G5); y la secundaria LP 16-395 (M4.0V). Sistema físico.
\item WDS 12406+4017: sistema cuádruple conformado por tres primarias: HD 110279 A (G:), HD 110279 B (G0) y TYC 3021-982-1 (F:); y la secundaria 2MASS J12403633+4017589 (M). En realidad, TYC 3021-982-1 no es una primaria, sino un candidato (descartado a ojo) de un sistema aparentemente cuádruple, posiblemente triple. Una imagen de este sistema puede verse en la Figura \ref{dossistemas}. 
\item WDS 13316+5857: sistema triple conformado por una primaria, HD 117845 (G2V) y dos secundarias: 2MASS J13313493+5857171 (M1) y PM I13312+5857 (M2.5). Sistema físico pero descartado porque el movimiento propio de la primaria es $<$50 mas a$^{-1}$.
\item 14575-2125: sistema triple conformado por una primaria, HD 131977 (K4V) y dos secundarias: HD 131976 (M1.5V) y GSC 06180-00916 (g:). Sistema óptico.
\item WDS 15282-0921: sistema triple conformado por dos primarias: HD 137763 (G9V) y HD 137778 (K2V); y una secundaria, G151-61 (M4.5). Sistema óptico.
\item WDS 19510+1025: sistema triple conformado por una primaria, $o$ Aql A (F8V); y dos secundarias: $o$ Aql B (M3.5V) y $o$ Aql C (M0.0V). Sistema físico.
\item WDS 21546-0318: sistema triple conformado por una primaria, HD 208177 (F5IV); y dos secundarias: BD-03 5329 B (m0:) y PM I21547-0318 (M3.5V). Sistema físico pero descartado porque el movimiento propio de la primaria es $<$50 mas a$^{-1}$.
\item WDS 23194+7900: sistema triple conformado por una primaria, V368 Cep A (G9V); y dos secundarias: V368 Cep B (M3.5V) y NLTT 56725 (M5.0V). Sistema físico pero la primaria es rotador rápido.   
\item WDS 23581+2420: sistema quíntuple conformado por tres primarias: HD 224459 (G5), BD+23 4830 B (G0) y BD+23 4830 C (G0); y dos secundarias: G 131-5 (M3) y G 131-6 (K7V). Sistema es físico pero la primaria es binaria espectroscópica (SB2) y el movimiento propio total de BD+23 4830 C $<$50 mas a$^{-1}$. Este sistema puede verse en la Figura~\ref{dossistemas}.

\end{itemize}

\noindent El resto de los sistemas de la muestra son binarios compuestos por una primaria FGK y una secundaria M o K tardía.\\

\subsubsection{Sistemas descartados}
\label{sist_desc}
\noindent Aplicando los criterios de selección mencionados en la Sección \ref{comprobacion}, se han descartado en total de 51 sistemas. El desglose puede verse en la Tabla \ref{desglose_descarte}. \\

\begin{table}[H]
\renewcommand{\tablename}{Tabla}
\caption{Número de sistemas descartados.}
\label{desglose_descarte}
\begin{center}
\begin{tabular}{l c}
\hline
\hline
\noalign{\smallskip}
Razón & Número\\

\noalign{\smallskip}
\hline
\noalign{\smallskip}

Movimientos propios diferentes & 10\\
Separación angular $\rho$ $<$ 5 arcsec & 16\\
Estrellas binarias espectroscópicas SB2 & 10\\
Movimiento propio primaria $<$ 50 mas a$^{-1}$ & 9\\
Primarias FGK rotadoras rápidas ($v \sin i>10$ km/s) & 4\\
Primarias muy calientes ($<$F4V) & 2\\
\hline
\noalign{\smallskip}
Total & 51\\

\noalign{\smallskip}
\hline

\end{tabular}
\end{center}
\end{table}

\noindent En el Apéndice \ref{aped.B} (Tabla \ref{descartados}) se detalla cada uno de los sistemas descartados y su razón. Mención importante al sistema WDS 04359+1631, conformado por Aldebarán A y B. Debido a su alta luminosidad, el espectro de la secundaria está muy contaminado y, por tanto, no es útil para introducirlo en $StePar$. Por otro lado, las estrellas HD 24916 B y HD 285970 B, en los sistemas WDS 03575-0110 y WDS 04429+1843 respectivamente, son binarias espectroscópicas que, aunque no se ha podido determinar su tipo (SB1, SB2...), se han descartado por precaución.

\subsubsection{Muestra final de estudio}
\noindent La muestra final de estudio, tras la limpieza explicada en la Sección \ref{sist_desc}, consta de 158 pares de estrellas, con 156 estrellas primarias y 156 secundarias, siendo los sistemas WDS 07041+7514, WDS 12406+4017, WDS 19510+1025 y WDS 23581+2420 los únicos múltiples en la muestra final. El par más lejano de la muestra final es WDS 08110+7955, situado a 200$\pm$43 pc de distancia, mientras que el más cercano es WDS 04153-0736, a una distancia de 4.9850$\pm$0.0057 pc. De los 158 pares, los sistemas WDS 00491+5749, WDS 03042+6142, WDS 04153-0736 y WDS 17050-0504 poseen un movimiento propio total $\mu$ mayor de 1000 mas a$^{-1}$. El par con el movimiento propio más alto de la muestra final es WDS 04153-0736 con un movimiento propio total de la estrella primaria, $o$ Eri, de $\mu=4088.57\pm0.21$ mas a$^{-1}$. Los cinco pares con el movimiento propio más alto de la muestra final pueden verse en la Tabla \ref{sistemas_rapidos}. Ahora se pueden realizar los gráficos de movimientos propios (Sección \ref{comprobacion}) para comprobar que, ahora sí, todos los sistemas son físicos. Estos gráficos se muestran en las Figuras \ref{pmraclean} y \ref{pmdeclean} en el Apéndice \ref{aped.A}. Se comprueba cómo ahora los puntos se disponen todos a lo largo de una recta, por lo que podemos afirmar que todos los sistemas de la muestra final son físicos y aptos para realizar las calibraciones de metalicidad. También puede verse de una manera más clara con los diagramas de las Figuras \ref{mua_mub} y \ref{mua_mub_limpio} (Apéndice \ref{aped.A}). En éstos, se representa la diferencia de movimientos propios totales de la componente primaria y secundaria frente al movimiento propio total de la primaria. Mientras que en la Figura \ref{mua_mub}, correspondiente a la muestra inicial, la dispersión es elevada, no lo es así en la Figura \ref{mua_mub_limpio} correspondiente a la muestra final. En efecto, aquellos puntos con mayor dispersión representan sistemas no físicos y, por ende, han sido descartados.\\

\begin{table}[H]
\renewcommand{\tablename}{Tabla}
\caption{Sistemas con el movimiento propio más alto.}
\label{sistemas_rapidos}
\begin{center}
\begin{tabular}{l l c c}
\hline
\hline
\noalign{\smallskip}
Identificador & Nombre primaria & $\mu_A$ & $\mu_B$\\
WDS & Simbad & [mas a$^{-1}$] & [mas a$^{-1}$]\\
\noalign{\smallskip}
\hline
\noalign{\smallskip}

04153-0736 & $o$ Eri A & 4088.57$\pm$0.21 & 4087.8$\pm$7.3\\
17050-0504 & HD 154363 A & 1461.48$\pm$0.89 & 1456.5$\pm$8.0\\
00491+5749 & $\eta$ Cas A & 1255.9$\pm$8.0 & 1209.8$\pm$2.5\\
03042+6142 & HD 18757 & 1001.50$\pm$0.60 & 1001.4$\pm$4.8\\
20036+2954 & HD 190360 A & 862.02$\pm$0.24 & 860.2$\pm$8.0\\

\noalign{\smallskip}
\hline

\end{tabular}
\end{center}
\end{table}

\subsection{Calibraciones de metalicidad}

\noindent Una vez se ha limpiado la lista de sistemas no aptos para la calibración, se pasa a continuación a describir las diferentes calibraciones de metalicidad que se han obtenido en este trabajo.

\subsubsection{Calibraciones espectroscópicas}

\noindent \underline{Muestra 1}: se han calculado valores de metalicidad para 41 secundarias con tipos espectrales desde K7V hasta M6V con las calibraciones de Rojas-Ayala et al. (2012), Terrien et al. (2012) y Newton et al. (2014). Estos valores se presentan en el Apéndice \ref{aped.B} (Tabla \ref{metalicidades_espectroscopicas}), al final del informe. El rango de anchuras equivalentes considerado para estos ajustes está entre 2.09 y 8.07 {\AA} para el doblete de Na~{\sc i} y entre 1.31 y 5.76 {\AA} para el triplete de Ca~{\sc i}, correspondiente a un intervalo de metalicidades de --0.88$<$[Fe/H]$<$+0.54. Los valores de metalicidad de secundarias obtenidos se han comparado con los valores de metalicidad de sus respectivas primarias obtenidos por el grupo de investigación de la UCM dedicado a CARMENES. Los coeficientes de regresión $R^{2}$ para dichos ajustes pueden verse en la Tabla \ref{muestra1_coef}.

\begin{table}[H]
\renewcommand{\tablename}{Tabla}
\caption{Coeficientes de regresión lineal.}
\label{muestra1_coef}
\begin{center}
\begin{tabular}{l c}
\hline
\hline
\noalign{\smallskip}
Metalicidad & Coeficiente de regresión $R^{2}$\\ 
\noalign{\smallskip}
\hline
\noalign{\smallskip}

Rojas-Ayala et al. (2012) & 0.79\\
Terrien et al. (2012) & 0.78\\
Newton et al. (2014) & 0.85\\

\noalign{\smallskip}
\hline

\end{tabular}
\end{center}
\end{table}

\noindent La afirmación de considerar iguales las metalicidades de primarias y secundarias en un mismo sistema es correcta debido a que los coeficientes de regresión son significativamente altos (hay que tener en cuenta que las medidas de anchuras equivalentes en estrellas M se ven afectadas por numerosas causas, como se ha explicado anteriormente). Los valores que mejor muestran esta relación son los obtenidos a partir de la calibración de Newton et al. (2014). Las gráficas de los ajustes se presentan en el Apéndice \ref{aped.A} (Figuras \ref{metra}, \ref{metter} y \ref{metnew}). \\

\noindent \underline{Muestra 2}: se han recopilado anchuras equivalentes de Na~{\sc i} para 89 secundarias con tipos espectrales desde K5V hasta M6V de Mann et al. (2013), Mann et al. (2014), Mann et al. (2015) y Newton et al. (2014). La calibración se ha realizado representando los valores de [Fe/H]$^{*}$ en función de las anchuras equivalentes del doblete de Na~{\sc i}. El mejor ajuste lo ofrece un polinomio de grado dos, como en la calibración derivada por Newton et al. (2014). La expresión de la calibración es:

\begin{equation}
\label{calibracionewton}
\text{[Fe/H]}=-1.61\,\text{dex}+0.48\,\text{EW(Na{\sc i})}-0.031\,\text{EW(Na{\sc i})}^{2}
\end{equation}

\begin{figure}[H]
\centering
\renewcommand{\figurename}{Figura}
\begin{subfigure}{0.9\textwidth}
  \centering
  \includegraphics[width=0.7\linewidth]{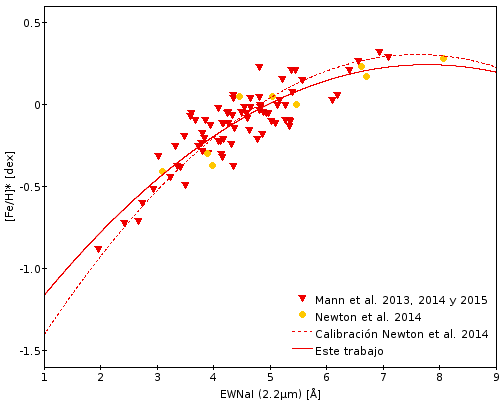}
\end{subfigure}
\caption {{\small Calibraciones de metalicidad obtenidas en este trabajo (línea continua) y Newton et al. (2014) (línea discontinua) para diferentes valores de anchuras equivalentes del doblete de Na~{\sc i}. Promedio de error en el eje X=0.17 \AA . Promedio de error en el eje Y=0.03 dex.}}
\label{mi_ajuste}
\end{figure}

\noindent Se ha calibrado para anchuras equivalentes de Na~{\sc i} entre 1.97 y 8.07 \AA, correspondiente a un rango de metalicidades de --0.79$<$[Fe/H]$<$+0.24. El coeficiente de correlación $R^{2}$ derivado de este ajuste es de 0.77, muy similar al obtenido por Newton et al. (2014) que es de 0.78. La Figura \ref{mi_ajuste} muestra estos dos ajustes.

\noindent La expresión \ref{calibracionewton} puede así servir para calcular la metalicidad de estrellas M aisladas conociendo únicamente la anchura equivalente del doblete de Na~{\sc i}, lo que facilita la identificación de las estrellas M más metálicas.\\

\subsubsection{Calibraciones fotométricas}

\noindent Como se ha explicado en la Subsección \ref{magfot}, el primer paso para la derivación de una calibración fotométrica es crear un contorno de isometalicidad en el plano \{\textit{$V-K_{S}$, M$_{K_S}$}\}. Para ello, se han seleccionado 32 estrellas primarias de la muestra con valores de metalicidad en un intervalo de 2$\sigma$ la metalicidad del Sol. El intervalo de metalicidades es --0.090$<$[Fe/H]$<$+0.060. Para estas estrellas, se han representado los valores de la magnitud absoluta en banda $K_{S}$ de sus correspondientes secundarias en función de su color $V-K_{S}$. No se ha encontrado magnitud $V$ para la estrella 2MASS J06364322+3751316, por lo que se ha excluído a su primaria, BD+37 1545, de la realización de esta parte, reduciéndose a 31 estrellas la derivación de la traza de isometalicidad. El mejor ajuste que describe este contorno de isometalicidad es un polinomio de tercer grado tal que $(V-K_{S})_{iso}=\Sigma a_{i} M_{K}^{i}$, donde los coeficientes del polinomio son: --13.7577, 7.4601, --1.0539, 0.0530 ($i$=0, 1, 2, 3). Esta traza de isometalicidad puede verse en la Figura \ref{ajuste_s2}, en el Apéndice \ref{aped.A}.\\

\begin{figure}[H]
\centering
\renewcommand{\figurename}{Figura}
\begin{subfigure}{1\textwidth}
  \centering
  \includegraphics[width=1\linewidth]{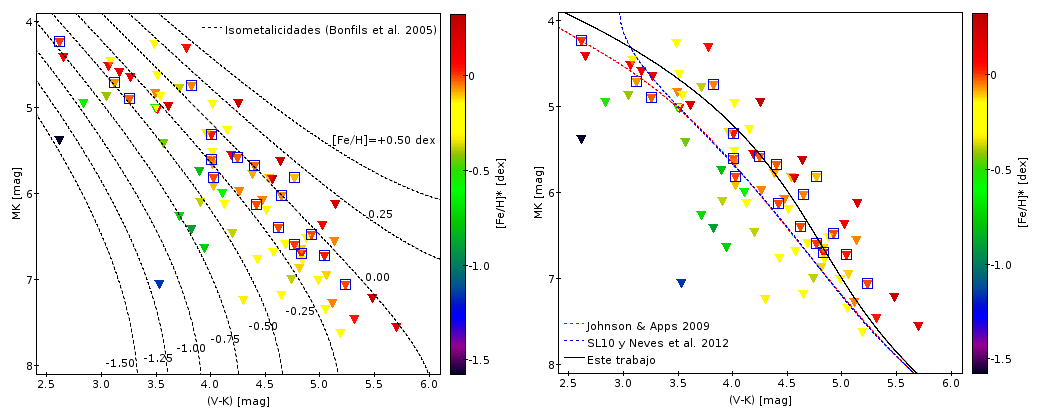}
\end{subfigure}
\caption {{\small Diagramas color-magnitud para las estrellas de la muestra final. Se han representado las isometalicidades de Bonfils et al. (2005) a la izquierda, y de Johnson \& Apps (2009; línea discontinua roja), Schlaufman \& Laughlin (2010; línea discontinua azul), Neves et al. (2012; línea discontinua azul) y este trabajo (línea continua negra) a la derecha. Promedio de error en el eje X=0.07 mag. Promedio de error en el eje Y=0.29 mag.}}
\label{calibrations}
\end{figure}

\noindent Para realizar la calibración fotométrica, se han tomado las 82 secundarias de la muestra final que poseen magnitud en banda $V$ de Zacharias et al. (2012) y que además tienen un valor de distancia y [Fe/H]$^{*}$. Estas estrellas pueden verse en la Tabla \ref{sigma2}, en el Apéndice \ref{aped.B}. La Figura \ref{calibrations} muestra estas estrellas en sendos diagramas color-magnitud con un código de colores que representan cuánto de metálicas son. Se han superpuesto sobre ellas las isometalicidades derivadas por Bonfils et al. (2005) en el panel de la izquierda y las correspondientes de Johnson \& Apps (2009), Schlaufman \& Laughlin (2010) y Neves et al. (2012) en el de la derecha, para sus respectivos valores de metalicidad media. En este último también se ha representado la traza de isometalicidad derivada en este trabajo. Los objetos recuadrados simbolizan aquellos cuya metalicidad es solar en un intervalo 2$\sigma$.

\noindent Como puede verse en la Figura \ref{calibrations}, existe una relación clara entre la metalicidad de la estrella y la posición que ésta ocupa en el diagrama color-magnitud, pues éstas parecen disponerse en planos diagonales al diagrama. Efectivamente, las estrellas más metálicas (rojas y amarillas en la Figura \ref{calibrations}) se sitúan en un plano diagonal superior a las menos metálicas (verdes y azules). Destacan los dos objetos con más baja metalicidad, situados en el plano inferior del gráfico: G 251--53 (WDS 08110+7955, [Fe/H]=--1.58 dex) y G 17--27 (WDS 16348--0412, [Fe/H]=--1.16 dex). Si se observan los contornos de isometalicidad de Bonfils et al. (2005), se ve cómo éstos no reproducen los valores de metalicidad solar, pues las estrellas con una metalicidad solar en un intervalo 2$\sigma$ se disponen, en su mayoría, entre las isometalicidades de [Fe/H]=0 y [Fe/H]=-0.50. Este hecho ya fue tenido en cuenta por Johnson \& Apps (2009) y posteriores. Estos autores afirmaron que la calibración de Bonfils et al. (2005) contenía un error sistemático de --0.32 dex para [Fe/H]$>$+0.2 dex. De aquí dedujeron que la relación obtenida por Bonfils et al. (2005) no era correcta, particularmente para [Fe/H]$>$0 dex. \\

\noindent Si ahora se comparan las trazas de Johnson \& Apps (2009), Schlaufman \& Laughlin (2010), Neves et al. (2012) y la derivada en este trabajo, se ve cómo las tres primeras (en realidad dos, pues Neves et al. (2012) toma el mismo polinomio de ajuste de Schlaufman \& Laughlin 2010) se sitúan por debajo de la aquí obtenida. Esto es debido a que, como se ha comentado anteriormente, las calibraciones de Johnson \& Apps, Schlaufman \& Laughlin y Neves toman la vecindad solar como isometálica (con una metalicidad media [Fe/H]$\sim$--0.12 dex), mientras que la calibración obtenida en este trabajo se ha hecho con estrellas de metalicidad solar en un intervalo 2$\sigma$ ([Fe/H]$\sim$0 dex). De esta manera, la traza de isometalicidad derivada en este trabajo caracteriza un espacio mucho más amplio que la propia vecindad solar.\\

\noindent La calibración de metalicidad se ha obtenido a partir de $\Delta(V-K_{S})$, es decir, la distancia entre el punto $V-K_{S}$ y la traza de isometalicidad que hemos hallado. Estos valores pueden verse en la Tabla \ref{ucac_ajuste} en el Apéndice \ref{aped.B}. La expresión que mejor ajusta estos valores de metalicidad es lineal en $\Delta(V-K_{S})$ y toma la siguiente forma funcional:

\begin{equation}
\label{mi_calibracion_foto}
\text{[Fe/H]}=0.58\Delta(V-K_{S})-0.07
\end{equation}

\noindent Esta ecuación es válida en el rango de 2.6 mag$<(V-K_{S})<$5.7 mag. La Tabla \ref{tabla_neves} es una adaptación de la publicada en Neves et al. (2012) y muestra otras calibraciones similares, además de la obtenida aquí, con el coeficiente de regresión lineal de cada estudio. Se comprueba que la calibración aquí determinada mejora las calibraciones previas deducidas por Schlaufman \& Laughlin (2010) y Neves et al. (2012), posiblemente debido a que esta muestra es considerablemente más amplia que previos estudios.

\begin{table}[H]
\renewcommand{\tablename}{Tabla}
\caption{Calibraciones fotométricas de varios autores junto su número de objetos de cada muestra y el coeficiente de regresión lineal para cada estudio (adaptación de Neves et al. 2012).}
\label{tabla_neves}
\begin{center}
\begin{tabular}{l c c}
\hline
\hline
\noalign{\smallskip}
Fuente de la calibración: ecuación & Nº de objetos de la muestra & $R^{2}$\\ 
\noalign{\smallskip}
\hline
\noalign{\smallskip}

B05: [Fe/H]=0.196-1.527$M_{K_{S}}$+0.091$M_{K_{S}}^{2}$+1.886$(V-K_{S})$-0.142$(V-K_S)^{2}$ & 20 & ...\\
JA09: [Fe/H]=0.56$\Delta M_{K_{S}}$-0.05, $\Delta M_{K_{S}}$=$MS-M_{K_{S}}$ & 6 & ...\\
SL10: [Fe/H]=0.79$\Delta(V-K_{S})$-0.17 & 19 & 0.49\\
N12: [Fe/H]=0.57$\Delta(V-K_{S})$-0.17 & 23 & 0.43\\
Este trabajo: [Fe/H]=0.58$\Delta(V-K_{S})$-0.07 & 82 & 0.57\\

\noalign{\smallskip}
\hline

\end{tabular}
\end{center}
\end{table}

\begin{figure}[H]
\centering
\renewcommand{\figurename}{Figura}
\begin{subfigure}{1\textwidth}
  \centering
  \includegraphics[width=0.57\linewidth]{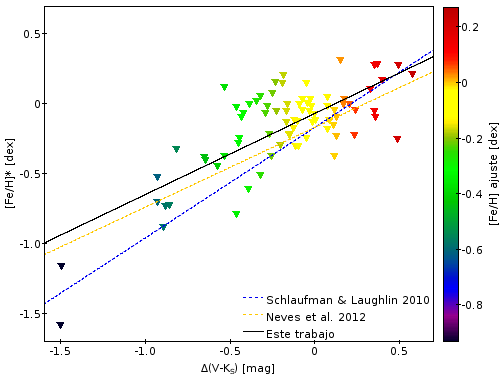}
\end{subfigure}
\caption {{\small Calibraciones de Schlaufman \& Laughlin (2010), Neves et al. (2012) y la obtenida en este trabajo. El código de colores representa las metalicidades calculadas con la calibración propia fotométrica. Promedio de error en el eje X=0.73 mag. Promedio de error en el eje Y=0.03 dex.}}
\label{avkvsfe}
\end{figure}

\noindent Para las 82 estrellas de este apartado, se han calculado valores de metalicidad según las calibraciones de la Tabla \ref{tabla_neves}. Estos resultados se muestran en la Tabla \ref{metalicidades_fotometricas}, en el Apéndice \ref{aped.B}. La Figura \ref{avkvsfe} muestra las calibraciones tanto de Schlaufman y Laughlin (2010) como la de Neves et al. (2012), además de la obtenida en este trabajo. Se ha representado en el eje X el factor $\Delta(V-K_{S})$ y en el Y, las metalicidades de calibración. El código de colores representa las metalicidades obtenidas en este trabajo según la Ecuación \ref{mi_calibracion_foto}.\\

\noindent La Figura \ref{avkvsfe} muestra que el término $\Delta(V-K_{S})$, es decir, la distancia a la traza isometálica, es un buen trazador de la metalicidad de las estrellas. También esta figura puede ayudar a comparar las metalicidades determinadas en este trabajo con las obtenidas por el grupo de investigación de la UCM, pues las primeras son cálculos fotométricos mientras que las segundas son espectroscópicas. En el régimen de bajas metalicidades, --1.5$<$[Fe/H]$<$--0.5, ambas parecen coincidir, pues el código de color morado/azul representa bien los puntos del diagrama. Una vez que se va aumentando en metalicidad, esta relación ya no es tan clara, ya que se observa algunas estrellas muy metálicas (rojas en el código de colores) que deberían tener una metalicidad menor según [Fe/H]$^{*}$. Esta dispersión puede ser debida a que las calibraciones usadas dependen fuertemente de la magnitud en banda $V$ de objetos muy débiles (8.7 mag $<V<$ 15.7 mag), por lo que los errores de fotometría de este tipo de estudios no son nada despreciables, aunque se aprecia la tendencia general de la metalicidad en la calibración.\\

\begin{figure}[H]
\centering
\renewcommand{\figurename}{Figura}
\begin{subfigure}{1\textwidth}
  \centering
  \includegraphics[width=0.85\linewidth]{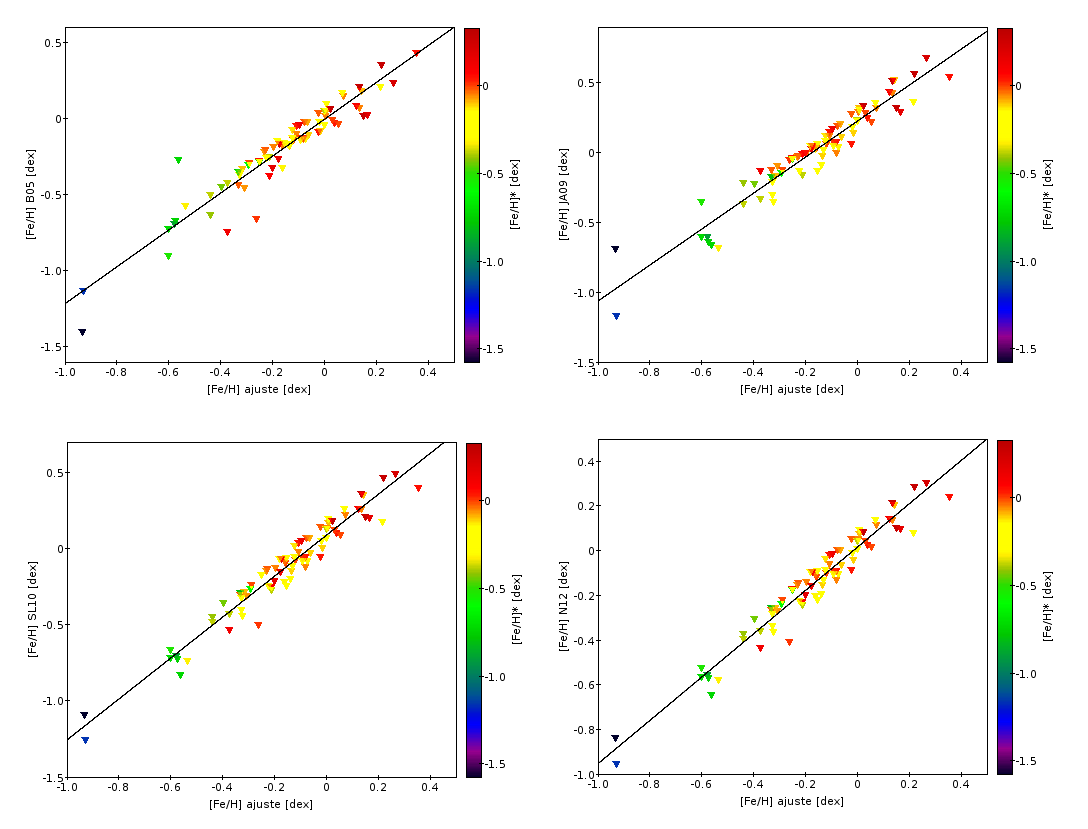}
\end{subfigure}
\caption {{\small Metalicidades determinadas por diferentes calibraciones en función de las calculadas en este trabajo. La línea negra representa el ajuste lineal de los puntos. Promedio de error en el eje X=0.42 dex. Promedio de error en el eje Y=0.17 dex.}}
\label{ajustes_todos}
\end{figure}

\noindent Para finalizar este trabajo, se han comparado las metalicidades calculadas con la calibración aquí obtenida con las correspondientes determinadas con las otras calibraciones de la Tabla \ref{tabla_neves}. La Figura \ref{ajustes_todos} muestra dicha comparación. En esta ocasión, el código de colores hace referencia a las metalicidades [Fe/H]$^{*}$. Los resultados del ajuste, junto con el rango de metalicidades hallado para cada calibración, se muestran en la Tabla \ref{metcal}.

\noindent Como se ve, los resultados obtenidos parecen ser bastante satisfactorios, pues todos ellos se disponen a lo largo de una recta claramente definida. Como cabría esperar, las dos calibraciones con mayor dispersión son las de Bonfils et al. (2005) y Johnson \& Apps (2009), ya que fueron las primeras y no consideran el factor $\Delta (V-K_{S})$ como trazador de metalicidad. Por contra, las calibraciones de Schlaufman \& Laughlin (2010) y Neves et al. (2012), que sí consideran este término como calibrador, son más similares a la aquí obtenida. Tanto en el panel de Bonfils et al. (arriba izquierda) como en los de Schlaufman \& Laughlin (abajo izquierda) y Neves et al. (abajo derecha) se encuentran dos estrellas (BD+44 4400 y TYC 3094-1728-1) muy metálicas (rojas) en una metalicidad que correspondería a [Fe/H]$\sim$ --0.30 dex. Resulta que estos dos objetos son de tipo espectral K5V, que, para esta calibración, están muy al límite de su aplicación, al ser demasiado calientes. De hecho, estos dos objetos también aparecen en la Figura \ref{calibrations} en la parte superior izquierda de ambos diagramas, reforzando la idea de que este tipo de calibraciones no son válidas para tipos espectrales tan tempranos ($\leq$ K5V). Aun así, se puede ver la tendencia anteriormente expuesta, situándose los objetos menos metálicos (azules según el análisis espectroscópico) en las partes inferiores del diagrama y enrojeciéndose cada vez más a medida que aumenta la metalicidad.\\

\begin{table}[H]
\renewcommand{\tablename}{Tabla}
\caption{Rango de metalicidades determinadas y comparación con las calculadas con la calibración obtenida en este trabajo.}
\label{metcal}
\begin{center}
\begin{tabular}{l c c}
\hline
\hline
\noalign{\smallskip}
Calibración & Rango de metalicidad & Coeficiente de regresión $R^{2}$\\ 
\noalign{\smallskip}
\hline
\noalign{\smallskip}

Bonfils et al. (2005) & --1.40$<$[Fe/H]$<$+0.43 & 0.90\\
Johnson \& Apps (2009) & --1.17$<$[Fe/H]$<$+0.68 & 0.92\\
Schlaufman \& Laughlin (2010) & --1.25$<$[Fe/H]$<$+0.49 & 0.94\\
Neves et al. (2012) & --0.95$<$[Fe/H]$<$+0.31 & 0.94\\
Este trabajo & --0.93$<$[Fe/H]$<$+0.36 & ... \\

\noalign{\smallskip}
\hline

\end{tabular}
\end{center}
\end{table}

\newpage
\section{Conclusiones}

\noindent El principal objetivo de este trabajo ha sido recopilar información de utilidad proveniente de varios catálogos de 209 pares estelares (muestra inicial) formados por una primaria de tipo espectral F, G o K y una secundaria de tipo M (o K tardía) como apoyo al grupo de investigación de la UCM dedicado a la ciencia preparatoria de CARMENES. Se han recopilados datos del propio sistema así como información astrométrica (coordenadas, paralajes y distancias), cinemática (movimientos propios), fotométrica (magnitudes en bandas $V$, $J$, $H$ y $K_{S}$) y espectroscópica (anchuras equivalentes de Na~{\sc i} y Ca~{\sc i} en banda $K$) de ambas componentes por separado para elaborar una calibración fotométrica y otra espectroscópica propias.\\

\noindent Con la ayuda de herramientas del observatorio virtual como $Aladin$ y $TopCat$, se han visualizado cada uno de los 209 sistemas de la muestra inicial y se han descartado aquellos que no cumplían los requisitos necesarios para realizar la calibración. Así, se han descartado 51 pares que, o no eran sistemas físicos (estudio de movimientos propios) o eran sistemas próximos o no cumplían otros criterios mencionados en el texto (binarias espectroscópicas SB2, rotadores rápidos...). De este modo, se ha elaborado una muestra final con 158 pares de estrellas (muestra final).\\

\noindent A partir de esta muestra final, se ha estudiado la metalicidad de las componentes M de los sistemas. Este proceso se ha realizado desde dos puntos de vista: espectroscopía y fotometría. En el estudio espectroscópico, se han calculado metalicidades a partir de diferentes calibraciones previas (Rojas-Ayala et al. 2012, Terrien et al. 2012 y Newton et al. 2014) y se ha derivado una  calibración propia dependiente únicamente de la anchura equivalente del doblete de Na~{\sc i} en banda $K$ de enanas M. La calibración se ha obtenido suponiendo una metalicidad de la primaria igual a la de la secundaria. Las metalicidades de la primaria han sido calculadas por el grupo de investigación de la UCM dedicado a CARMENES.\\

\noindent Por otro lado, se han calculado metalicidades a partir de los datos fotométricos de la secundaria siguiendo las calibraciones de Bonfils et al. (2005), Johnson \& Apps (2009), Schlaufman \& Laughlin (2010) y Neves et al. (2012) y se ha obtenido una calibración fotométrica propia tomando como traza de isometalicidad aquellas estrellas primarias de la muestra final cuyo valor de metalicidad estaba dentro de 2$\sigma$ la metalicidad del Sol. De esta forma, esta calibración no sólo es válida para la vecindad solar sino que puede aplicarse hasta una distancia mayor.\\

\noindent En total, se han estimado metalicidades, espectroscópicas y/o fotométricas, para 134 enanas M con un rango de metalicidades --0.9$<$[Fe/H]$<$+0.4, y se han comparado estos valores con los calculados a partir de otras calibraciones, obteniendo una muy buena relación entre ellos.\\

\section*{Trabajo futuro}
\label{futuro}
\addcontentsline{toc}{section}{Futuro del proyecto}

\noindent En relación al futuro y al grupo de investigación de la UCM dedicado a CARMENES, la muestra final resultado de este TFM permitirá terminar alguno de los trabajos ya iniciados sobre calibraciones de la metalicidad con los índices espectroscópicos de espectros de baja resolución de enanas M (espectros CAFOS), que culminará con una publicación que está ahora en preparación (Alonso-Floriano et al., in prep.). A su vez, permitirá perfeccionar las calibraciones derivadas de este TFM con datos adicionales y fotometría en otros colores como AllWISE ($W1-W2$) (Montes et al. 2012, 2016 y Alonso-Floriano et al., in prep.). Así, se espera que toda esta información pueda ser aplicada a la base de datos de CARMENES, Carmencita, con el objetivo de poder estudiar la relación entre la metalicidad y la presencia de planetas de diferentes tipos en este rango de masas (estrellas M) y comprobar si se comporta de igual manera que en las estrellas mas masivas (FGK).\\

\newpage
\section*{Agradecimientos}
\addcontentsline{toc}{section}{Agradecimientos}
\noindent La realización de este Trabajo Fin de Máster no hubiera sido posible sin el apoyo de algunas personas que me gustaría mencionar a continuación. \\
\noindent En primer lugar, agradecer a mis dos tutores, los doctores David Montes y José Antonio Caballero, la dedicación que me han brindado. A pesar de lo apretado de sus agendas, siempre han hecho un hueco para atender y resolver mis dudas. A David Montes, por haberme proporcionado las referencias necesarias para realizar este trabajo y por la confianza depositada en mí a la hora de incluir mis gráficas en sus pósters. Es un orgullo ver mi nombre publicado al lado de algunos de los astrofísicos más prestigiosos del mundo. A José Antonio Caballero, por haberme enseñado la investigación de verdad, por mostrarme el trabajo que hay detrás de ella, por las horas y horas reunidos en un pequeño despacho de la UCM observando sistemas en $Aladin$... incluso por las charlas de música que teníamos en los descansos. \\
\noindent Quiero también mostrar toda mi gratitud al grupo de investigación de la UCM dedicado a CARMENES, en especial a Miriam, Hugo y Javier. Aparte de acogerme en su lugar de trabajo, siempre se han mostrado abiertos y dispuestos a ayudarme con mis dudas. Gracias por enseñarme la importancia del trabajo en equipo. Mención especial para Javier, el cual ha sido una de las "columnas" sobre las que se apoya este trabajo. Nunca tuve ningún problema en entrar en su despacho y preguntarle sobre tal o cual sistema. Él ha sido el enlace entre el mundo real y el de la investigación, aconsejándome y escuchándome siempre que lo he necesitado. \\
\noindent Por extensión, gracias al consorcio CARMENES y a algunos nombres que reconoceré allá por donde vaya (Andreas Quirrenbach, Pedro J. Amado...) por su labor científica. Espero que vuestra investigación obtenga resultados que supongan un salto cualitativo en el conocimiento de la Astrofísica planetaria. Ha sido un placer formar parte, aunque sólo hayan sido unos meses, de algo tan importante como CARMENES.\\

\newpage

\section*{Referencias}
\addcontentsline{toc}{section}{Referencias}
\noindent \medskip
\noindent Abt, H. A. 2009, ApJ, 180, A117\\ \medskip
\noindent Alonso-Floriano, F. J., Morales, J. C., Caballero, J. A. et al. 2015, A\&A, 577, A128\\ \medskip
\noindent Alonso-Floriano, F. J., Caballero, J. A., Cortés-Contreras, M. et al. 2015, A\&A, 583, A85\\ \medskip
\noindent Alonso Floriano, F. J. 2016, PhD Tesis, Universidad Complutense de Madrid, España\\ \medskip
\noindent Alonso-Floriano, F. J., Montes, D. Tabernero, H. M. et al. 2016, SEA (Bilbao) Póster\\ \medskip
\noindent Anglada-Escudé, G., Amado, P. J., Barnes, J. et al. 2016, Nature, 536, 437\\ \medskip
\noindent Bhatia, A. S. 2005, ed. New Delhi: Deep \& Deep Publications\\ \medskip
\noindent Boeshaar, P. C. 1976, PhD Thesis, Ohio State Univ., Columbus, EE. UU.\\ \medskip
\noindent Boeshaar, P. C. \& Tyson, J. A. 1985, AJ, 90, 817\\ \medskip
\noindent Bonfils, X., Delfosse, X., Udry, S., Santos, N. C. 2005, A\&A, 442, 635\\ \medskip
\noindent Caballero, J. A. 2006, PhD Tesis, Universidad de La Laguna, España\\ \medskip 
\noindent Caballero, J. A. 2009, A\&A, 507, 251\\ \medskip 
\noindent Caballero, J. A. 2010, A\&A, 514, A98\\ \medskip
\noindent Caballero, J. A., Cortés-Contreras, M., López-Santiago, J. et al. 2013, Highlights of Spanish Astrophysics~{\sc vii}, 645\\ \medskip
\noindent Caballero, J. A., Cortés-Contreras, M, Alonso-Floriano, F. J. et al. 2016, Cool Stars 19, Uppsala (Suecia).\\ \medskip
\noindent Carleo, I., Sanna, N., Gratton, R. et al. 2016, ExA, 41, 351\\ \medskip
\noindent de Bruijne, J. H. J., Eilers, A. C. 2012, A\&A, 546, A61\\ \medskip
\noindent Dommanget, J. \& Nys, O. 2000, A\&A, 363, 991\\ \medskip
\noindent Donati, J. F., Delfosse, X., Artigau, E. et al. 2014, SPIROU-2000-IRAP-RP-00503\\ \medskip
\noindent Finch, C. T., Zacharias, N. 2016, yCat, 1333, 0F\\ \medskip
\noindent Frith, J., Pinfield, D. J., Jones, H. R. A. et al. 2013, MNRAS, 435, 2161\\ \medskip
\noindent Gagné, J., Plavchan, P., Gao, P. et al. 2016, ApJ, 822, 40\\ \medskip
\noindent Goldberg, D., Mazeh, T., Latham, D. W. 2003, ApJ, 591, 1\\ \medskip
\noindent Gould, A. \& Chanamé, J. 2004, ApJ, 150, 455\\ \medskip
\noindent H$\o$g, E., Fabricius, C., Makarov, V.V. et al. 2000, A\&A, 355, 27\\ \medskip
\noindent Ivanov, G.A. 2008, KFNT, 24, 480\\ \medskip
\noindent Johnson, J. A., Apps, K. 2009, ApJ, 699, 933\\ \medskip
\noindent Kharchenko, N.V., Roeser, S. et al. 2001, KFNT, 17, 409\\ \medskip
\noindent Kirkpatrick, J. D., Henry, T. J. \& McCarthy, D. W., Jr. 1991, ApJS, 77, 417\\ \medskip
\noindent Klemola A. R., Hanson R. B., Jones, B. F. 1987, AJ, 94, 501\\ \medskip
\noindent Lasker, B., Lattanzi M. G., McLean B. J. et al. 2008, ApJ, 136, 735\\ \medskip
\noindent Lazorenko, P. et al. 2015, yCat, 1332, 0\\ \medskip
\noindent Lee, S., Yuk, I., Lee, H. et al. 2010, SPIE, 7735E, 2KL\\ \medskip
\noindent Lepine, S. et al. 2011, AJ, 142, 138\\ \medskip
\noindent Lurie, J. C., Henry, T. J., Jao, W. C. et al. 2014, AJ, 148, 91\\ \medskip
\noindent Mahadevan, S., Ramsey, L. W., Terrien, R. et al. 2014, SPIE, 9147, 1GM\\ \medskip
\noindent Maiolino, R., Haehnelt, M., Murphy, M. T. et al. 2013, arXiv1310, 3163M\\ \medskip
\noindent Mann, A. W., Brewer, J. M., Gaidos, E. et al. 2013, AJ, 145, 52\\ \medskip
\noindent Mann, A. W., Deacon, N. R., Gaidos, E. et al. 2014, AJ, 147, 160\\ \medskip
\noindent Mann, A. W., Feiden, G. A., Gaidos, E. et al. 2015, ApJ, 804, 64\\ \medskip
\noindent Martín, E. L., Guenther, E., Zapatero Osorio, M. R. 2006, ApJ, 644, 75\\ \medskip
\noindent Mason, B. D., Wycoff, G. L., Hartkopf, W. I. et al. 2001, 2001, AJ, 122, 3466\\ \medskip
\noindent Mayor, M. \& Queloz, D. 1995, Nature, 378, 355\\ \medskip
\noindent Montes, D., Alonso-Floriano, F. J., Caballero, J. A. et al. (2012) Cool Stars 17 (Barcelona) Póster\\ \medskip
\noindent Montes, D., Alonso-Floriano, F. J., Caballero, J. A. et al. (2016) RIA Space Tec (Madrid) Póster\\ \medskip
\noindent Montes, D., Alonso-Floriano, F. J., Tabernero, H. M. et al. (2016) Cool Stars 19 (Uppsala) Póster\\ \medskip
\noindent Neves, V., Bonfils, X., Santos, N. C. 2012, A\&A, 538, A25\\ \medskip
\noindent Newton, E. R., Charbonneau, D., Irwin, J. et al. 2014, AJ, 147, 20\\ \medskip
\noindent Pasinetti-Fracassini, L. E. et al. 2001, A\&A, 367, 521\\ \medskip
\noindent Perryman, M. A. C., Lindegren, L., Kovalevsky, J. et al. 1997, A\&A, 323, 49\\ \medskip 
\noindent Phillips, N. M., Greaves, J. S., Dent, W. R. F. et al. 2010, MNRAS, 403, 1089\\ \medskip
\noindent Qi, Z. X., Yu, Y., Bucciasrelli, B. et al. 2015, yCat, 1331, 0Q\\ \medskip
\noindent Quirrenbach, A., Amado, P. J., Caballero, J. A. et al. 2014, SPIE, 9147, E1F\\ \medskip
\noindent Raskin, G., Van Winckel, H., Hensenberge, H. et al. 2011, A\&A, 526, A69\\ \medskip
\noindent Reid, I. N. \& Cruz, K.L. 2002, AJ, 123, 2806\\ \medskip
\noindent Reid, I. N. et al. 2004, AJ, 128, 463\\ \medskip
\noindent Reid, I. N. \& Hawley, S. L. 2005, Praxis Publishing Ltd, Chichester, Reino Unido\\ \medskip
\noindent Ren, S., Fu, Y. 2010, AJ, 139, 1975\\ \medskip 
\noindent Roeser, S., Demleitner, M., \& Schilbach, E. 2010, AJ, 139, 2440\\ \medskip
\noindent Rojas-Ayala, B., Covey, K. R., Muirhead, P. S. et al. 2010, ApJ, 720, 113\\ \medskip
\noindent Rojas-Ayala, B., Covey, K. R., Muirhead, P. S. et al. 2012, ApJ, 748, 93\\ \medskip
\noindent Salim, S. \& Gould A. 2003, ApJ, 582, 1011\\ \medskip
\noindent Schlaufman, K. C. \& Laughlin, G. 2010, A\&A, 519, A105\\ \medskip
\noindent Seifahrt, A. \& Käufl, H. U. 2008, A\&A, 491, 929\\ \medskip
\noindent Skidmore, W. 2015, RAA, 15, 1945S\\ \medskip
\noindent Skiff, B. A. 2009, yCat, 102023\\ \medskip
\noindent Skrutskie, M. F., Cutri, R. M., Stiening, R. et al. 2006, AJ, 131, 1163\\ \medskip
\noindent Smart, R. L., Nicastro, L. 2013, yCat, 1324, 0S\\ \medskip
\noindent Soubiran, C. et al. 2016, A\&A, 591, A118\\ \medskip
\noindent Tabernero, H. M., González Hernández, J. I., Montes, D. 2013, Highlights of Spanish Astrophysics~{\sc vii}, 673\\ \medskip
\noindent Tabernero, H. M., Montes, D., González Hernández, J. I. 2012, A\&A, 547, A13\\ \medskip
\noindent Tamura, M., Suto, H., Nishikawa, J. et al. 2012, SPIE, 8446, 1TT\\ \medskip
\noindent Terrien, R. C., Mahadevan, S., Bender, C. F. et al. 2012, ApJ, 747, 38\\ \medskip
\noindent Terrien, R.C., Mahadevan, S., Bender C.F. et al. 2015, ApJ, 802, 10\\ \medskip
\noindent Tokovinin, A. 2014, AJ, 147, 86\\ \medskip
\noindent Valenti, J. A. \& Fischer, D. A. 2005, ApJS, 159, 141\\ \medskip
\noindent van Altena, W. F., Lee J. T., Hoffleit, E. D. et al. 1995, GCTP, C, 0V\\ \medskip
\noindent van Leeuwen, F. 2007, A\&A, 474, 653\\ \medskip
\noindent Zacharias N., Monet D. G., Levine S. E. et al. 2005, AAS, 205, 4815\\ \medskip
\noindent Zacharias N., Finch C. T., Girard T. M. et al. 2012, yCat, 1322, 0\\ \medskip

\newpage

\renewcommand\appendixpagename{{\LARGE Apéndices}}

\appendixpage
\renewcommand{\appendixname}{Apendice}
\renewcommand\thefigure{\thesection.\arabic{figure}} 
\begin{appendices}

\section{Figuras}
\label{aped.A}
\setcounter{figure}{0}

\noindent A continuación se adjuntan las figuras adicionales que no se han incluido en el texto por falta de espacio. En primer lugar, se presentan gráficas de comparación de movimientos propios pertenecientes a las muestras inicial y final. A continuación, se muestran las comparaciones de las metalicidades espectroscópicas [Fe/H]$^{*}$ con las calculadas a partir de las calibraciones de Rojas-Ayala et al. (2012), Terrien et al. (2012) y Newton et al. (2014), respectivamente. Por último, se adjunta la traza de isometalicidad utilizada para la elaboración de la calibración fotométrica.\\

\begin{figure}[H]
\centering
\renewcommand{\figurename}{Figura}
\begin{subfigure}{1\textwidth}
  \centering
  \includegraphics[width=0.60\linewidth]{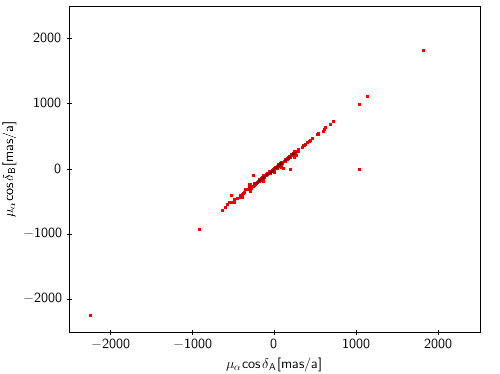}
\end{subfigure}
\caption {{\small Movimientos propios en ascensión recta de la primaria frente a los de la secundaria para la muestra inicial. Promedio de error en el eje X=1.18 mas a$^{-1}$. Promedio de error en el eje Y=6.00 mas a$^{-1}$.}}
\label{fig6}
\end{figure}

\begin{figure}[H]
\centering
\renewcommand{\figurename}{Figura}
\begin{subfigure}{1\textwidth}
  \centering
  \includegraphics[width=0.60\linewidth]{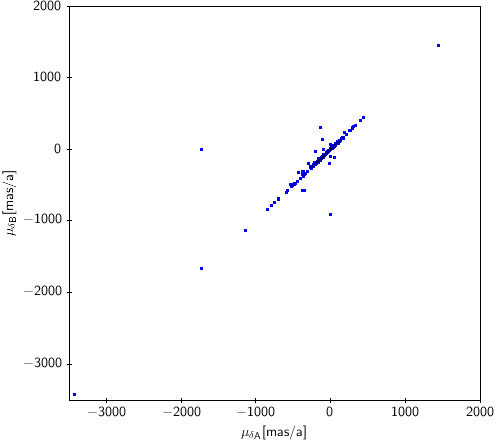}
\end{subfigure}
\caption {{\small Movimientos propios en declinación de la primaria frente a los de la secundaria para la muestra inicial. Promedio de error en el eje X=0.95 mas a$^{-1}$. Promedio de error en el eje Y=5.94 mas a$^{-1}$.}}
\label{fig7}
\end{figure}

\begin{figure}[H]
\centering
\renewcommand{\figurename}{Figura}
\begin{subfigure}{1\textwidth}
  \centering
  \includegraphics[width=0.7\linewidth]{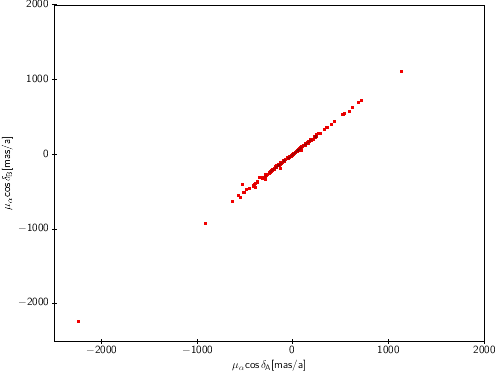}
\end{subfigure}
\caption {{\small Movimientos propios en ascensión recta de la primaria frente a los de la secundaria para la muestra final. Promedio de error en el eje X=1.23 mas a$^{-1}$. Promedio de error en el eje Y=6.18 mas a$^{-1}$.}}
\label{pmraclean}
\end{figure}

\begin{figure}[H]
\centering
\renewcommand{\figurename}{Figura}
\begin{subfigure}{1\textwidth}
  \centering
  \includegraphics[width=0.7\linewidth]{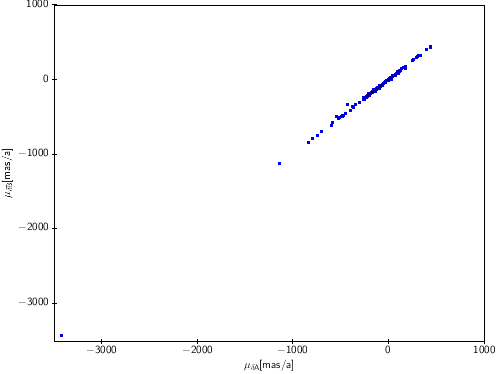}
\end{subfigure}
\caption {{\small Movimientos propios en declinación de la primaria frente a los de la secundaria para la muestra final. Promedio de error en el eje X=0.97 mas a$^{-1}$. Promedio de error en el eje Y=6.15 mas a$^{-1}$.}}
\label{pmdeclean}
\end{figure}

\begin{figure}[H]
\centering
\renewcommand{\figurename}{Figura}
\begin{subfigure}{1\textwidth}
  \centering
  \includegraphics[width=0.7\linewidth]{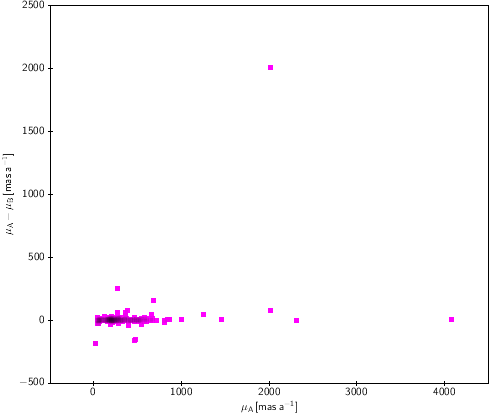}
\end{subfigure}
\caption {{\small Diferencia de movimientos propios totales de la componente primaria y secundaria frente al movimiento propio total de la primaria para la muestra inicial. Promedio de error en el eje X=1.07 mas a$^{-1}$. Promedio de error en el eje Y=6.21 mas a$^{-1}$.}}
\label{mua_mub}
\end{figure}

\begin{figure}[H]
\centering
\renewcommand{\figurename}{Figura}
\begin{subfigure}{1\textwidth}
  \centering
  \includegraphics[width=0.7\linewidth]{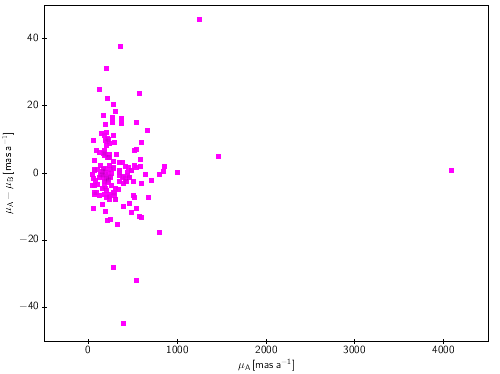}
\end{subfigure}
\caption {{\small Diferencia de movimientos propios totales de la componente primaria y secundaria frente al movimiento propio total de la primaria para la muestra final. Promedio de error en el eje X=1.10 mas a$^{-1}$. Promedio de error en el eje Y=6.41 mas a$^{-1}$. Notese la diferencia de escala en el eje vertical}}
\label{mua_mub_limpio}
\end{figure}

\begin{figure}[H]
\centering
\renewcommand{\figurename}{Figura}
\begin{subfigure}{1\textwidth}
  \centering
  \includegraphics[width=0.7\linewidth]{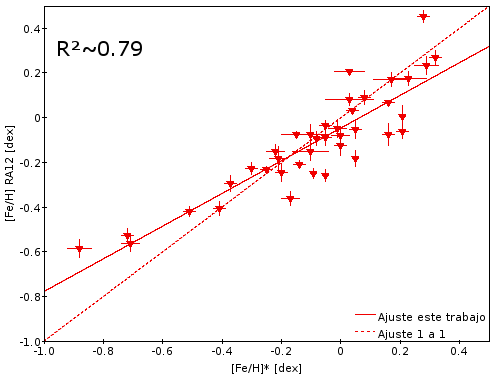}
\end{subfigure}
\caption {{\small Metalicidades espectroscópicas calculadas a partir de la calibración de Rojas-Ayala et al. (2012) frente a las metalicidades [Fe/H]$^{*}$.}}
\label{metra}
\end{figure}

\begin{figure}[H]
\centering
\renewcommand{\figurename}{Figura}
\begin{subfigure}{1\textwidth}
  \centering
  \includegraphics[width=0.7\linewidth]{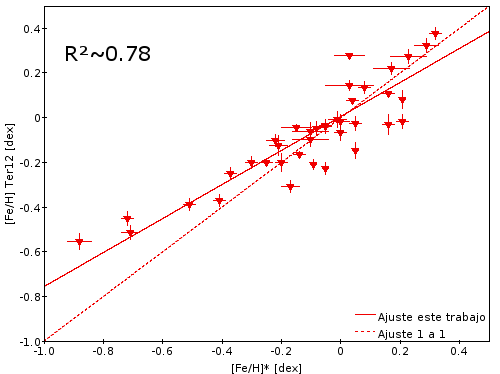}
\end{subfigure}
\caption {{\small Metalicidades espectroscópicas calculadas a partir de la calibración de Terrien et al. (2012) frente a las metalicidades [Fe/H]$^{*}$.}}
\label{metter}
\end{figure}

\begin{figure}[H]
\centering
\renewcommand{\figurename}{Figura}
\begin{subfigure}{1\textwidth}
  \centering
  \includegraphics[width=0.7\linewidth]{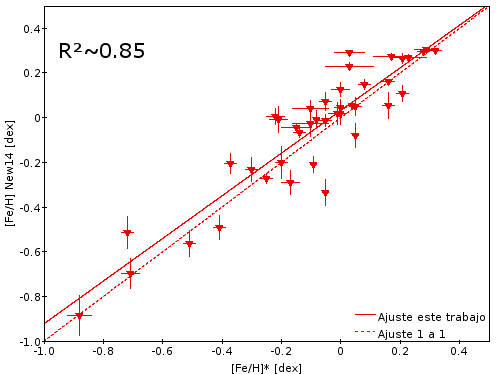}
\end{subfigure}
\caption {{\small Metalicidades espectroscópicas calculadas a partir de la calibración de Newton et al. (2014) frente a las metalicidades [Fe/H]$^{*}$.}}
\label{metnew}
\end{figure}

\begin{figure}[H]
\centering
\renewcommand{\figurename}{Figura}
\begin{subfigure}{1\textwidth}
  \centering
  \includegraphics[width=0.7\linewidth]{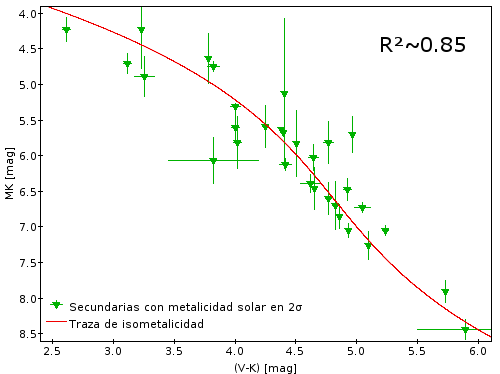}
\end{subfigure}
\caption {{\small Secundarias con metalicidad solar en 2$\sigma$ y la traza de isometalicidad (línea roja).}}
\label{ajuste_s2}
\end{figure}

\newpage

\section{Tablas} 
\label{aped.B}
\renewcommand\thetable{\thesection.\arabic{table}} 
\setcounter{table}{0}  

\noindent A continuación se adjuntan las tablas adicionales que no se han incluido en el texto por falta de espacio. En primer lugar, se muestra la muestra inicial de sistemas que se han estudiado en este TFM. Le siguen las tablas de astrometría y cinemática (información sobre paralajes, movimientos propios...) y fotometría. A continuación, la tabla con los pares desechados y una breve explicación, para pasar a las tablas con la información espectroscópica (Muestras 1 y 2) y fotométrica utilizada para los cálculos. Las últimas dos tablas corresponden a las metalicidades, tanto espectroscópicas como fotométricas, determinadas en este trabajo.\\

\begin{center}
\begin{ThreePartTable}
\begin{TableNotes}\footnotesize
\item[a] Referencias de código de descubridor -- BAK: Bakos, G.A.; BKO: Berko, E.; BU: Burnham, S.W.; BVD: Benavides, R; CAB: Caballero, J.A.; COU: Couteau, P.; DAM: Damm, F.; DON: Donner, H.F.; EIS: Eisenbeiss, T.; ENG: Engelmann, R; ES: Espin, T.E.; GAL: Gallo, J; GIC: Giclas et al.; GJ: Gliese, W. \& Jahreiss, H.; GRV: Greaves, J.; GWP: Garraf Wide Pairs; H: Herschel, W.; HJ: Herschel, J.F.W; HO: Hough, G.W.; HU: Hussey, W.J.; HZG: Hertzsprung, E.; J: Jonckheere, R.; JNN: Janson, M.; KUI: Kuiper, G.P.; LAF: Lafreniere, D.; LDS: Luyten, W. J.; LEP: Lepine, S. et al.; LMP: Lampens, P.; PLQ: Paloque, E.; RAG: Raghavan, D. et al.; S: South, J.; SKF: Skiff, B.A; SLE: Soulie, G.; STF: Struve, F. J. W.; STN: Stone, O.;  STT: Struve, O;  TOK: Tokovinin, A.A.; UC: USNO CCD Astrographic Catalog (UCAC1, UCAC2, UCAC3, UCAC4); VBS:Van Biesbroeck, G.
\item[b] Las estrellas con nombre en formato JXXXXXXXX$\pm$XXXXXXX son estrellas del catálogo 2MASS (Skrutskie et al. 2006).
\end{TableNotes}
\renewcommand{\tablename}{Tabla}
\begin{longtable}{l c c c l c c c}
\caption{Muestra inicial de pares.}
\label{muestracompleta}\\
\hline
\hline
\noalign{\smallskip}
Identificador & Código \tnote{a} & $\rho$ & $\theta$ & Nombre \tnote{b} & $\alpha$ & $\delta$ & Tipo \\
WDS & descubridor & [arcsec] & [deg] & Simbad & (J2000) & (J2000) & espectral \\
\noalign{\smallskip}
\hline
\noalign{\smallskip}
\endfirsthead

\hline
\insertTableNotes
\endlastfoot

00153+5304 &  &  &  & G 217-41 & 00:15:14.8 & +53:04:27 & K3\\
 & GIC 5 & 18.8 & 355 & G 217-40 & 00:15:14.6 & +53:04:46 & M2.5V\\
 \noalign{\smallskip}
00385+4300 & & & & BD+42 126  & 00:38:29.2  & +43:00:00 & G5\\
 & LDS 5176 & 53.1 & 125 & LP 193-345 & 00:38:33.2 & +42:59:30 & M0.5V\\
 \noalign{\smallskip}
00452+0015 &  &  &  & HD 4271 A & 00:45:11.0 & +00:15:12 & F8\\
 & LDS 836 & 55.4 & 45 & HD 4271 B & 00:45:13.6 & +00:15:51 & M4.0V\\
 \noalign{\smallskip}
00467-0426 &  &  &  & HD 4449 & 00:46:40.5 & -04:25:37 & G5V\\
 & LDS 9100 & 66.9 & 40 & LP 646-9 & 00:46:43.4 & -04:24:46 & M4.0V\\
\end{longtable}
\end{ThreePartTable}
\end{center}

\newpage

\begin{landscape}
\begin{center}
\begin{ThreePartTable}
\begin{TableNotes}\footnotesize
\item[a] Referencias de paralajes--[1]: de Bruijne et al. (2012); [2]: Gould et al. (2004); [3]: Kharchenko et al. (2009); [4]: Lurie et al. (2014); [5]: Perryman et al. (1997); [6]: Reid et al. (2002); [7]: Reid et al. (2004); [8]: Tokovinin (2014); [9]: van Altena et al. (1995); [10]: van Leeuwen et al. (2007);
\item [b] Referencias de movimientos propios--[1]: Caballero (2009); [2]: Finch et al. (2016); [3]: Frith et al. (2013); [4]: Gould et al. (2004); [5]: H$\o$g et al. (2000); [6]: Ivanov (2008); [7]: Phillips et al. (2010); [8]: Qi et al. (2015); [9]: Roeser et al. (2010); [10]: Salim \& Gould(2003); [11]: Smart \& Nicastro (2013); [12]: Terrien et al. (2015); [13]: van Altena et al. (1995); [14]: van Leuween et al. (2007); [15]: Zacharias et al. (2005); [16]: Zacharias et al. (2012); [17]: Zacharias et al. (2015) [17].\\
Nota: las estrellas con [*] en movimientos propios tenían referencia Zacharias et al. (2012) pero se les ha asignado otros valores más precisos.
\end{TableNotes}

\renewcommand{\tablename}{Tabla}
\begin{longtable}{l l c c c c c c c c}
\caption{Astrometría de la muestra inicial.} 
\label{astrometry}\\
\hline
\hline
\noalign{\smallskip}
Identificador & Nombre & $\pi$ \tnote{a}& $d$ & $\mu_\alpha \cos{\delta}$ \tnote{b}& $\mu_{\delta}$ & $\mu$ & $\dfrac{\Delta\mu_\alpha \cos{\delta}}{\overline{\mu_\alpha \cos{\delta}}}$ & $\dfrac{\Delta\mu_{\delta}}{\overline{\mu_{\delta}}}$ & $\dfrac{\Delta\mu}{\mu_{A}}$\\
WDS & Simbad & [mas] & [pc] & [mas a$^{-1}$] & [mas a$^{-1}$] & [mas a$^{-1}$] &  &  & \\
\noalign{\smallskip}
\hline
\noalign{\smallskip}
\endfirsthead

\hline
\insertTableNotes
\endlastfoot

00153+5304 & G 217-41 & 10.7$\pm$2.3 $^{[10]}$& 94$\pm$20 & 224.2$\pm$1.9 $^{[14]}$& 46.3$\pm$1.6 & 228.9$\pm$1.9 & 0.011& 0.051& 0.014\\
 & G 217-40 & 12.2$\pm$2.1 $^{[2]}$& 82$\pm$14 & 226.6$\pm$7.0 $^{[16]}$& 44.0$\pm$2.7 & 230.8$\pm$6.9\\
\noalign{\smallskip}
00385+4300 & BD+42 126 & 20.2$\pm$1.8 $^{[10]}$& 49.6$\pm$4.3 & 187.2$\pm$1.1 $^{[14]}$& -81.83$\pm$0.93 & 204.3$\pm$1.1 & 0.036 & 0.035 & 0.036\\
 & LP 193-345 & 19.9$\pm$1.9 $^{[2]}$& 50.4$\pm$4.7 & 194.0$\pm$8.0 $^{[16]}$& -79.0$\pm$8.0 & 209.5$\pm$8.0\\
\noalign{\smallskip}
00452+0015 & HD 4271 A & 24.17$\pm$0.55 $^{[10]}$& 41.37$\pm$0.94 & 266.23$\pm$0.55 $^{[14]}$& -51.30$\pm$0.37 & 271.13$\pm$0.54 & 0.25 & 0.22 & 0.22\\
 & HD 4271 B & 24.27$\pm$0.88 $^{[2]}$& 41.2$\pm$1.5 & 207.0$\pm$8.0 $^{[16]}$& -41.0$\pm$8.0 & 211.0$\pm$8.0\\
\noalign{\smallskip}
00467-0426 & HD 4449 & 34.77$\pm$2.73 $^{[10]}$& 28.8$\pm$2.3 & 29.1$\pm$2.4 $^{[14]}$& -269.7$\pm$1.6 & 271.2$\pm$1.6 & 0.33 & 0.031 & 0.043\\
 & LP 646-9 & 35.7$\pm$2.7 $^{[2]}$& 28.0$\pm$2.1 & 20.9$\pm$8.0 $^{[16]}$& -261.4$\pm$8.0 & 262.2$\pm$8.0\\
\end{longtable}
\end{ThreePartTable}
\end{center}
\end{landscape}

\newpage

\begin{center}
\begin{ThreePartTable}
\begin{TableNotes}\footnotesize
\item[a] Referencias de magnitud $V$-- Dom00: Dommanget \& Nys (2000); ESA97: The Hipparcos and Tycho Catalogues (1997); GSC2.3: Lasker et al. (2008); Gou04: Gould \& Chaname (2004); Kle87: Klemola et al. (1987); Laz15: Lazorenko et al. (2015); Lep11: Lepine et al. (2011); Man13: Mann et al. (2003); NOMAD: Zacharias et al. (2005); Pas01: Pasinetti-Fracassini et al. 2001; RA12: Rojas Ayala (2012); Rei04: Reid et al. (2004); Sal03: Salim \& Gould (2003); Sou16: Soubiran et al. (2016); Tok14: Tokovinin et al. (2014); UCAC4: Zacharias et al. (2012).
\item[b] Magnitudes $J$, $H$ y $K$ obtenidas de 2MASS.
\end{TableNotes}

\renewcommand{\tablename}{Tabla}
\begin{longtable}{l l c c c c c}
\caption{Fotometría de la muestra inicial.} 
\label{photometry}\\
\hline
\hline
\noalign{\smallskip}
Identificador & Nombre & $V$ & Ref.\tnote{a} & $J$ \tnote{b} & $H$ \tnote{b} & $K_{s}$ \tnote{b}\\
WDS & Simbad & [mag] & & [mag] & [mag] & [mag]\\
\noalign{\smallskip}
\hline
\noalign{\smallskip}
\endfirsthead

\hline
\insertTableNotes
\endlastfoot

00153+5304 & G 217-41 & 10.71$\pm$0.99 & UCAC4 & 8.838$\pm$0.024 & 8.440$\pm$0.018 & 8.372$\pm$0.046\\
 & G 217-40 & 14.39$\pm$0.99 & Rei04 & 10.823$\pm$0.023 & 10.231$\pm$0.022 & 9.986$\pm$0.018\\
 \noalign{\smallskip}
00385+4300 & BD+42 126 & 9.934$\pm$0.030 & UCAC4 & 8.411$\pm$0.020 & 7.934$\pm$0.021 & 7.850$\pm$0.027\\
 & LP 193-345 & 13.449$\pm$0.030 & UCAC4 & 10.489$\pm$0.023 & 9.936$\pm$0.021 & 9.736$\pm$0.019\\
 \noalign{\smallskip}
00452+0015 & HD 4271 A & 8.732$\pm$0.050 & UCAC4 & 5.984$\pm$0.019 & 5.694$\pm$0.020 & 5.630$\pm$0.023\\
 & HD 4271 B & 14 & Tok14 & 10.114$\pm$0.027 & 9.561$\pm$0.023 & 9.264$\pm$0.021\\  
 \noalign{\smallskip}
00467-0426 & HD 4449 & 9.054$\pm$0.010 & UCAC4 & 6.093$\pm$0.020 & 5.703$\pm$0.038 & 5.583$\pm$0.023\\
  & LP 646-9 & 15.096$\pm$0.010 & UCAC4 & 11.198$\pm$0.026 & 10.585$\pm$0.025 & 10.279$\pm$0.023\\
\end{longtable}
\end{ThreePartTable}
\end{center}

\newpage

\begin{center}
\begin{ThreePartTable}
\begin{TableNotes}\footnotesize
\item[a] Referencias al código adjunto -- BAG: Balega, Y.Y.; CAT: Catala, C. et al.; COU: Couteau, P.; ENG: Engelmann, R; GKI: Golimowski, D.A.; KUI: Kuiper, G.P.; LSC: Lowell-Southern Connecticut; RAO: Riddle, R.L. et al.; RST: Rossiter, R.A.; STF: Struve, F.G.W.  
\end{TableNotes}
\renewcommand{\tablename}{Tabla}
\begin{longtable}{l l l}
\caption{Sistemas descartados.}
\label{descartados}\\
\hline
\hline
\noalign{\smallskip}
Identificador & Nombre & Razón \tnote{a}\\
WDS & & \\
\noalign{\smallskip}
\hline
\noalign{\smallskip}
\endfirsthead

\hline
\insertTableNotes
\endlastfoot

00452+0015 & HD 4271 A & Secundaria es binaria visual próxima $\rho$=1.5 arcsec (RAO 4).\\
 & HD 4271 B & \\
 \noalign{\smallskip}
00467-0426 & HD 4449 & Primaria es binaria visual próxima $\rho$=0.10 arcsec (LSC 10).\\
 & LP 646-9 & \\
 \noalign{\smallskip}
01187-0052 & HD 7895 A & Primaria es binaria visual próxima $\rho$=1.0 arcsec (ENG 1).\\
 & HD 7895 B & \\
 \noalign{\smallskip}
01230-1258 & HD 8389 A & Secundaria es binaria visual próxima $\rho$=1.6 arcsec (KUI 6).\\
 & HD 8389 B & \\
\end{longtable}
\end{ThreePartTable}
\end{center}

\newpage

\begin{center}
\begin{ThreePartTable}
\begin{TableNotes}\footnotesize
\item[a] Código de referencias -- New14: Newton et al. (2014); Roj10: Rojas-Ayala et al. (2010); Roj12: Rojas-Ayala et al. (2012).
\end{TableNotes}

\renewcommand{\tablename}{Tabla}
\begin{longtable}{l l c c c c}
\caption{Espectroscopía en banda $K$ de secundarias compañeras (Muestra 1).} 
\label{muestra1}\\
\hline
\hline
\noalign{\smallskip}
Identificador & Nombre & EW(Na~{\sc i}) & EW(Ca~{\sc i}) & H$_2$O--$K$ & Ref.\tnote{a}\\
WDS & & [\AA] & [\AA] & & \\
\noalign{\smallskip}
\hline
\noalign{\smallskip}
\endfirsthead

\hline
\insertTableNotes
\endlastfoot

00153+5304 & G 217-40 & 7.006$\pm$0.055 & 3.954$\pm$0.055 & 0.8940$\pm$0.0020 & Roj12\\
01076+2257 & HD 6660 B & 6.19$\pm$0.16 & 3.63$\pm$0.14 & 0.8940$\pm$0.0050 & New14\\
02482+2704 & LP 354-414 & 5.47$\pm$0.19 & 2.68$\pm$0.18 & 0.8580$\pm$0.0050 & New14\\
03042+6142 & G 246-22 & 3.90$\pm$0.18 & 3.29$\pm$0.17 & 0.9240$\pm$0.0050 & New14\\
03150+0101 & BD+00 549 B & 2.09$\pm$0.21 & 1.98$\pm$0.27 & 0.9420$\pm$0.0050 & New14\\
03480+4032 & UCAC4 653-017808 & 8.07$\pm$0.15 & 5.76$\pm$0.15 & 0.9580$\pm$0.0050 & New14\\
04153-0736 & $o^{02}$ Eri C & 3.99$\pm$0.16 & 2.13$\pm$0.21 & 0.8350$\pm$0.0050 & New14\\
04559+0440 & HD 31412 B & 5.60$\pm$0.15 & 4.84$\pm$0.20 & 0.9650$\pm$0.0050 & New14\\
\end{longtable}
\end{ThreePartTable}
\end{center}

\newpage

\begin{center}
\begin{ThreePartTable}
\begin{TableNotes}\footnotesize
\item[a] Código de referencias -- New14: Newton et al. (2014); Man13: Mann et al. (2013); Man14: Mann et al. (2014); Man15: Mann et al. (2015).
\end{TableNotes}

\renewcommand{\tablename}{Tabla}
\begin{longtable}{l l c c}
\caption{Espectroscopía en banda $K$ de secundarias compañeras (Muestra 2).} 
\label{muestra2}\\
\hline
\hline
\noalign{\smallskip}
Identificador & Nombre & EW(Na~{\sc i}) & Ref.\tnote{a}\\
WDS & & [\AA] & \\
\noalign{\smallskip}
\hline
\noalign{\smallskip}
\endfirsthead

\hline
\insertTableNotes
\endlastfoot

00153+5304 & G 217-40 & 5.2 & Man13\\
01055+1526 & HD 6440 B & 3.3 & Man13\\
01076+2257 & HD 6660 B & 6.1 & Man13\\
01226+1245 & BD+12 168 B & 4.8 & Man13\\
01572-1015 & HD 11964 B & 4.4 & Man13\\
02291+2252 & BD+22 353 B & 3.6 & Man13\\
02442+4914 & $\theta$ Per B & 4.4 & Man13\\
02482+2704 & LP 354-414 & 5.5 & New14\\
\end{longtable}
\end{ThreePartTable}
\end{center}

\newpage

\begin{center}
\begin{ThreePartTable}
\begin{TableNotes}\footnotesize
\item[a] Todos los valores obtenidos con $StePar$ (Tabernero et al. 2012).
\end{TableNotes}

\renewcommand{\tablename}{Tabla}
\begin{longtable}{l l c c c c c}
\caption{Primarias con [Fe/H] solar en 2$\sigma$ \tnote{a}} 
\label{sigma2}\\
\hline
\hline
\noalign{\smallskip}
Identificador & Nombre & SpT & \textit{T$_{\rm eff}$} & $log$ $g$ & $\xi$ & [Fe/H]$^{*}$\\
WDS & & & [K] & & [km/s] & [dex]\\
\noalign{\smallskip}
\hline
\noalign{\smallskip}
\endfirsthead

\hline
\insertTableNotes
\endlastfoot

00153+5304 & G 217-41 & K3 & 5160$\pm$73 & 4.43$\pm$0.17 & 0.59$\pm$0.20 & 0.030$\pm$0.051\\
01076+2257 & HD 6660 A & K4V & 4789$\pm$160 & 4.38$\pm$0.41 & 0.79$\pm$0.43 & 0.030$\pm$0.081\\
01215+3120 & EN Psc & K2V & 5052$\pm$50 & 4.42$\pm$0.14 & 0.75$\pm$0.16 & -0.020$\pm$0.032\\
01226+1245 & BD+12 168 A & K3V & 4911$\pm$140 & 4.33$\pm$0.31 & 0.65$\pm$0.44 & -0.030$\pm$0.091\\
02291+2252 & BD+22 353 A & K7V & 5181$\pm$58 & 4.40$\pm$0.15 & 0.53$\pm$0.18 & -0.070$\pm$0.041\\ 
02442+4914 & $\theta$ Per A & F8V & 6389$\pm$38 & 4.560$\pm$0.080 & 1.670$\pm$0.050 & 0.040$\pm$0.032\\  
02482+2704 & BC Ari & K1V & 5298$\pm$35 & 4.410$\pm$0.090 & 0.960$\pm$0.080 & 0.000$\pm$0.032\\ 
05003+2508 & HD 31867 A & G2V & 5677$\pm$23 & 4.450$\pm$0.070 & 0.850$\pm$0.040 & -0.010$\pm$0.022\\
\end{longtable}
\end{ThreePartTable}
\end{center}

\newpage

\begin{center}

\renewcommand{\tablename}{Tabla}
\begin{longtable}{l l c c c c c}
\caption{Secundarias seleccionadas para el ajuste fotométrico.} 
\label{ucac_ajuste}\\
\hline
\hline
\noalign{\smallskip}
Identificador & Nombre & $(V-K)$ & $M_{K}$ & $\Delta(V-K)$ & & [Fe/H]$^{*}$\\
WDS & & [mag] & [mag] & [mag] &  & [dex] \\
\noalign{\smallskip}
\hline
\noalign{\smallskip}
\endfirsthead

\hline
\endlastfoot

00385+4300 & LP 193-345 & 3.713$\pm$0.036 & 6.26$\pm$0.44 & -0.94$\pm$0.22 & & -0.700$\pm$0.081\\
01055+1526 & HD 6440 B & 3.482$\pm$0.017 & 4.26$\pm$0.19 & 0.42$\pm$0.24 & & -0.250$\pm$0.032\\
01340-0141 & LP 588-9 & 3.94$\pm$0.038 & 6.63$\pm$0.71 & -0.89$\pm$0.34 & & -0.730$\pm$0.032\\
01450-0104 & LP 588-44 & 4.20$\pm$0.14 & 6.45$\pm$0.42 & -0.55$\pm$0.25 & & -0.370$\pm$0.032\\
01572-1015 & HD 11964 B & 3.502$\pm$0.040 & 5.01$\pm$0.10 & -0.33$\pm$0.10 & & 0.060$\pm$0.022\\ 
03042+6142 & G 246-22 & 4.513$\pm$0.071 & 6.183$\pm$0.080 & -0.104$\pm$0.086 & & -0.300$\pm$0.022\\  
03078+2533 & HD 19381 B & 5.026$\pm$0.023 & 6.36$\pm$0.31 & 0.32$\pm$0.16 & & 0.110$\pm$0.022\\ 
03150+0101 & BD+00 549 B & 3.824$\pm$0.046 & 6.40$\pm$0.72 & -0.90$\pm$0.36 & & -0.880$\pm$0.041\\
\end{longtable}
\end{center}

\newpage

\begin{center}
\begin{ThreePartTable}
\begin{TableNotes}\footnotesize
\item[a] Valores proporcionados por $StePar$ (Tabernero et al. 2012). No han sido calculados en este trabajo.
\end{TableNotes}

\renewcommand{\tablename}{Tabla}
\begin{longtable}{l l c c c c c}
\caption{Metalicidades espectroscópicas (Muestras 1 y 2) calculadas en esta memoria.} 
\label{metalicidades_espectroscopicas}\\
\hline
\hline
\noalign{\smallskip}

Identificador & Nombre & [Fe/H]$^{*}$ \tnote{a} & RA12 & Ter12 & New14 & Este trabajo\\
WDS & Simbad & [dex] & [dex] & [dex] & [dex] & [dex]\\

\noalign{\smallskip}
\hline
\noalign{\smallskip}
\endfirsthead

\hline
\insertTableNotes

\endlastfoot

00153+5304 & G 217-40 & 0.030$\pm$0.050 & 0.2082$\pm$0.0097 & 0.2767$\pm$0.0086 & 0.2915$\pm$0.0027 & 0.03\\
01055+1526 & HD 6440 B & -0.250$\pm$0.030 & ... & ... & ... & -0.37\\
01076+2257 & HD 6660 B & 0.030$\pm$0.080 & 0.081$\pm$0.026 & 0.142$\pm$0.024 & 0.227$\pm$0.018 & 0.16\\
01226+1245 & BD+12 168 B & -0.030$\pm$0.090 & ... & ... & ... & -0.03\\
01572-1015 & HD 11964 B & 0.060$\pm$0.020 & ... & ... & ... & -0.12\\
02291+2252 & BD+22 353 B & -0.070$\pm$0.040 & ... & ... & ... & -0.29\\
02442+4914 & $\theta$ Per B & 0.040$\pm$0.030 & ... & ... & ... & -0.12\\
02482+2704 & LP 354-414 & 0.000$\pm$0.030 & -0.081$\pm$0.033 & -0.017$\pm$0.029 & 0.127$\pm$0.032 & 0.08\\
\end{longtable}
\end{ThreePartTable}
\end{center}

\newpage

\begin{center}

\renewcommand{\tablename}{Tabla}
\begin{longtable}{l l c c c c c}
\caption{Metalicidades fotométricas calculadas en esta memoria.} 
\label{metalicidades_fotometricas}\\
\hline
\hline
\noalign{\smallskip}

Identificador & Nombre & BO05 & JA09 & SL10 & N12 & Este trabajo\\
WDS & Simbad & [dex] & [dex] & [dex] & [dex] & [dex]\\

\noalign{\smallskip}
\hline
\noalign{\smallskip}
\endfirsthead

\hline
\endlastfoot

00385+4300 & LP 193-345 & -0.72$\pm$0.17 & -0.60$\pm$0.25 & -0.72$\pm$0.23 & -0.57$\pm$0.16 & -0.60$\pm$0.57\\
01055+1526 & HD 6440 B & 0.21$\pm$0.15 & 0.36$\pm$0.11 & 0.178$\pm$0.063 & 0.081$\pm$0.045 & 0.22$\pm$0.28\\
01340-0141 & LP 588-9 & -0.67$\pm$0.23 & -0.64$\pm$0.40 & -0.72$\pm$0.35 & -0.57$\pm$0.25 & -0.57$\pm$0.95\\
01450-0104 & LP 588-44 & -0.42$\pm$0.18 & -0.33$\pm$0.26 & -0.43$\pm$0.24 & -0.36$\pm$0.17 & -0.37$\pm$0.56\\
01572-1015 & HD 11964 B & -0.276$\pm$0.072 & -0.037$\pm$0.063 & -0.168$\pm$0.068 & -0.168$\pm$0.049 & -0.25$\pm$0.14\\
03042+6142 & G 246-22 & -0.123$\pm$0.051 & 0.088$\pm$0.072 & -0.051$\pm$0.066 & -0.084$\pm$0.047 & -0.12$\pm$0.11\\
03078+2533 & HD 19381 B & 0.08$\pm$0.12 & 0.44$\pm$0.18 & 0.26$\pm$0.16 & 0.14$\pm$0.11 & 0.12$\pm$0.41\\
03150+0101 & BD+00 549 B & -0.69$\pm$0.26 & -0.60$\pm$0.40 & -0.70$\pm$0.36 & -0.56$\pm$0.26 & -0.58$\pm$0.94\\
\end{longtable}
\end{center}

\end{appendices}

\newpage

\section*{Pósters}
\addcontentsline{toc}{section}{Pósters}

\noindent Aquí se adjuntan las publicaciones en formato Póster presentados en el congreso Cool Stars 19 Uppsala (Proceedings from the 19th Cambridge Workshop on Cool Stars, Stellar Systems, and the Sun, hosted by Uppsala University in Uppsala, Sweden from 06 – 10 June 2016, Edited by G. A. Feiden, \url{https://zenodo.org/collection/user-cs19}) y en el congreso SEA 2016 (XII Reunión Científica de la Sociedad Española de Astronomía (SEA) en Bilbao entre los días 18 y 22 de julio de 2016, 
Highlights of Astronomy and Astrophysics IX, \url{http://www.sea-astronomia.es/drupal/SEA2016}), respectivamente, que han sido de utilidad y han servido de referencia para la realización de este trabajo. Este TFM ha contribuido a las secciones de la calibración espectroscópica a partir de la anchura equivalente del doblete de Na~{\sc i} (2.2$\mu$m) y la calibración fotométrica a partir del diagrama color-magnitud con la realización de sendas gráficas.\\

\newgeometry{left=0.5cm,right=0cm,top=0.5cm,bottom=0cm}

\begin{center}
\begin{figure}[H] 
\center{\includegraphics[scale=0.23]{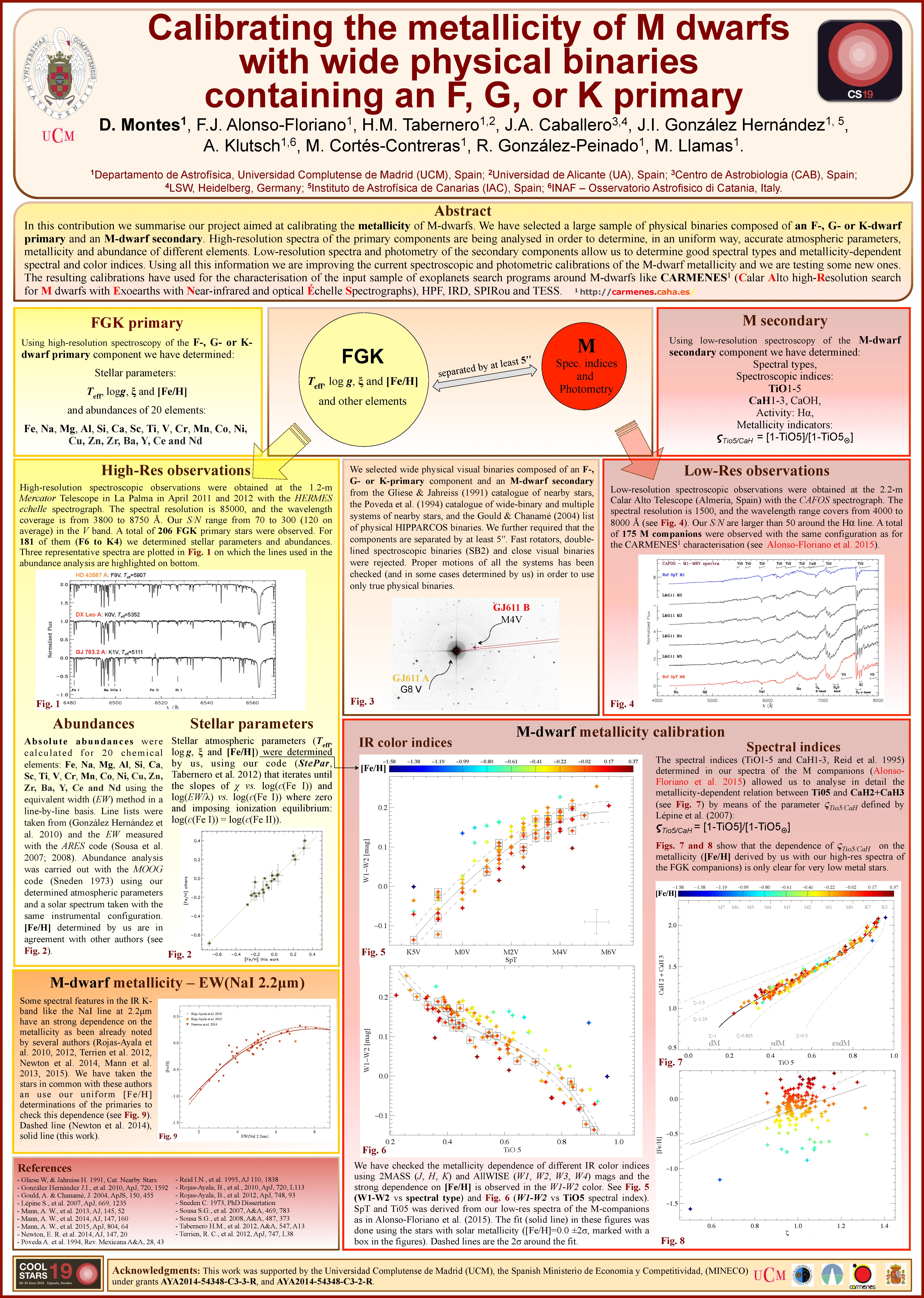}}
\label{Poster}
\end{figure}
\end{center}

\newpage
\begin{center}
\begin{figure}[H] 
\center{\includegraphics[scale=0.24]{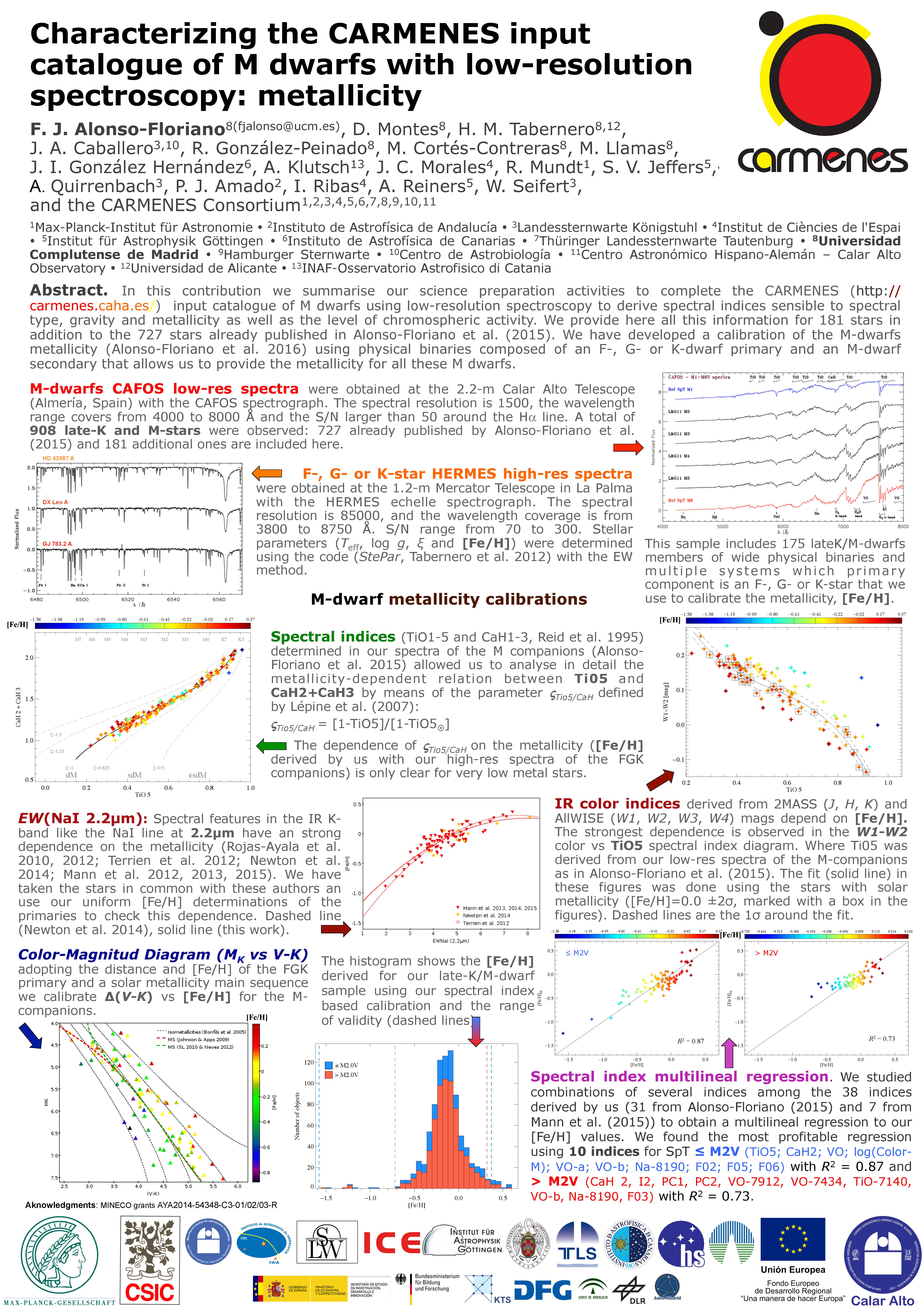}}
\label{Poster2}
\end{figure}
\end{center}

\restoregeometry

\end{document}